\documentclass[sort, compress,
3p,
11pt,
review
]{elsarticle}
\usepackage[utf8]{inputenc}
\usepackage[T1]{fontenc}
\usepackage{graphicx}
\usepackage{longtable}
\usepackage{wrapfig}
\usepackage{rotating}
\usepackage[normalem]{ulem}
\usepackage{amsmath}
\usepackage{amssymb}
\usepackage{capt-of}
\usepackage{lipsum}  
\usepackage{tikz}
\usepackage{textcomp}
\usepackage{natbib}
\usepackage{braket}
\usepackage{nicefrac}
\usepackage{bm}
\usepackage[dvipsnames]{xcolor}
\usetikzlibrary{shapes.geometric, arrows, positioning, calc}
\usepackage[hidelinks]{hyperref}
\usepackage{threeparttable}
\usepackage{lineno}
\usepackage{caption}         
\usepackage{amsmath} % or simply amstext

\DeclareRobustCommand{\circledcustom}[3]{%
  \tikz[baseline=(c.base)]{
    \node[draw, circle, inner sep=#2, outer sep=0pt] (c)
         {\raisebox{#1}{#3}};
  }%
}

\hypersetup{
 colorlinks=false,
 pdfauthor={Francesco Ricci},
 pdftitle={WOT-SRSH Automated Workflow},
 pdfkeywords={},
 pdfsubject={},
 pdflang={English}}

\begin{document}
\author[aff1]{Stephen E. Gant\corref{contrib}}
\author[aff2,aff3,aff4]{Francesco Ricci\corref{contrib}}
\author[aff5]{Guy Ohad}
%\author[aff1]{Sijia Ke}
\author[aff6,aff7]{Ashwin Ramasubramaniam}
\author[aff5]{Leeor Kronik}
\author[aff1,aff2,aff8]{Jeffrey B. Neaton\corref{corauthor}}

% Affiliations
\address[aff1]{Department of Physics, University of California Berkeley, Berkeley CA, United States}
\address[aff2]{Materials Sciences Division, Lawrence Berkeley National Laboratory, Berkeley, CA 94720, United States}
\address[aff3]{Université catholique de Louvain (UCLouvain), Institute of Condensed Matter and Nanosciences (IMCN), B-1348 Louvain-la-Neuve, Belgium}
\address[aff4]{Matgenix SRL, A6K Advanced Engineering Centre, 6000 Charleroi, Belgium}
\address[aff5]{Department of Molecular Chemistry and Materials Science, Weizmann Institute of Science, Rehovoth 76100, Israel}
\address[aff6]{Department of Mechanical and Industrial Engineering, University of Massachusetts Amherst, Amherst, MA 01003, United States}
\address[aff7]{Materials Science and Engineering Graduate Program, University of Massachusetts Amherst, Amherst, MA 01003, Unites States}
\address[aff8]{Kavli Energy NanoScience Institute at Berkeley, Berkeley, United States}

%\date{\today}
\title{Automated Workflow for Non-Empirical Wannier-Localized Optimal Tuning of Range-Separated Hybrid Functionals}
\cortext[contrib]{Both authors contributed equally}
\cortext[corauthor]{jbneaton@lbl.gov}
%\tableofcontents
\begin{abstract}
We introduce an automated workflow for generating non-empirical Wannier-localized optimally-tuned screened range-separated hybrid (WOT-SRSH) functionals.
WOT-SRSH functionals have been shown to yield highly accurate fundamental band gaps, band structures, and optical spectra for bulk and 2D semiconductors and insulators.
%Previously, WOT‑SRSH functionals have been manually tuned using physical intuition and multiple hybrid DFT supercell calculations.
Our workflow automatically and efficiently determines the WOT-SRSH functional parameters for a given crystal structure and composition, approximately enforcing the correct screened long-range Coulomb interaction and an ionization potential ansatz.
In contrast to previous manual tuning approaches, our tuning procedure relies on a new search algorithm that only requires a few hybrid functional calculations with minimal user input.
We demonstrate our workflow on 23 previously studied semiconductors and insulators, reporting the same high level of accuracy.
By automating the tuning process and improving its computational efficiency, the approach outlined here enables applications of the WOT-SRSH functional to compute spectroscopic and optoelectronic properties for a wide range of materials.
\end{abstract}

\maketitle

\section{Introduction}
Understanding and predicting spectroscopic properties of crystals is central to the discovery of new materials for electronic and optoelectronic applications.
Of particular importance is the calculation of the fundamental band gap, a key property that governs many electronic and optical phenomena, that can be defined as the difference between the ionization potential and the electron affinity of the material.
Developing first-principles methods capable of reliably screening tens of thousands of compounds' spectroscopic properties remains a significant challenge.
High-throughput calculations of band gaps and optical spectra require workflows that are accurate, non-empirical, efficient, and automated so that they can run unattended on high-performance computers.
Meeting these requirements has become a major bottleneck for integrating electronic-structure tools into both small-scale studies and inverse-design pipelines.
Even for a single material, the manual tuning and convergence testing of ``gold-standard'' \textit{ab initio} methods can consume a large amount of time and resources.
Chief among these methods are \textit{ab initio} many-body perturbation theory techniques such as the use of the $GW$ approximation for predicting band gaps~\cite{hybertsen_firstprinciples_1985, hybertsen_electron_1986, onida_electronic_2002} and the Bethe-Salpeter equation (BSE) approach for calculating neutral excitations~\cite{rohlfing_electronhole_1998, rohlfing_electronhole_2000, onida_electronic_2002}.

%I hvae commented out the old 2 paragraphs and put below a first attempt at a simplification/merge
A lower-cost alternative with competitive accuracy is density-functional theory (DFT) for band gaps and time-dependent DFT (TDDFT) for optical excitations~\cite{onida_electronic_2002}.
However, achieving quantitative accuracy within DFT has been a significant challenge~\cite{onida_electronic_2002, kummel_orbitaldependent_2008, borlido_exchangecorrelation_2020, bryenton_delocalization_2023}.
Even with the exact (and unknown) exchange-correlation functional, the Kohn-Sham (KS) eigenvalue gap generally differs from the fundamental gap~\cite{perdew_physical_1983,sham_densityfunctional_1983} by the so-called derivative discontinuity error~\cite{ullrich_chapter_2016,perdew_densityfunctional_1982}.
Guided by this insight, a wide family of approaches have been devised to restore---or at least approximate---the derivative discontinuity, improving gap predictions across bulk, low-dimensional materials~\cite{bylander_good_1990, geller_computational_2001, heyd_energy_2005, cococcioni_linear_2005, anisimov_transition_2005, ferreira_approximation_2008, shimazaki_band_2008, zhao_calculation_2009, tran_accurate_2009, chan_efficient_2010, dabo_firstprinciples_2010, dabo_koopmans_2010, marques_densitybased_2011, refaely-abramson_gap_2013, skone_selfconsistent_2014, borghi_koopmanscompliant_2014, gorling_exchangecorrelation_2015, skone_nonempirical_2016, ma_using_2016, weng_wannier_2017, perdew_understanding_2017, verma_hle17_2017, nguyen_koopmanscompliant_2018, cui_doubly_2018, chen_nonempirical_2018, miceli_nonempirical_2018, colonna_screening_2018, bischoff_nonempirical_2019, aschebrock_ultranonlocality_2019, colonna_koopmanscompliant_2019, weng_wannier_2020, cipriano_band_2020, tancogne-dejean_parameterfree_2020, lee_firstprinciples_2020, lorke_koopmanscompliant_2020, wing_band_2021, colonna_koopmans_2022, yang_oneshot_2022, yang_rangeseparated_2023, zhan_nonempirical_2023, degennaro_blochs_2022, ohad_optical_2023, linscott_koopmans_2023, camarasa-gomez_excitations_2024, schubert_predicting_2024, zhan_dielectricdependent_2025}.
Among these approaches, the generalized Kohn-Sham (gKS) scheme~\cite{seidl_generalized_1996, baer_timedependent_2018,garrick_exact_2020} stands out as a rigorous DFT framework that can, in principle, absorb the missing derivative discontinuity and improve the accuracy of computed band gaps.
Among gKS approaches, screened range-separated hybrids (SRSHs)~\cite{refaely-abramson_gap_2013, refaely-abramson_solidstate_2015, chen_nonempirical_2018} have shown great promise, and the recently proposed fully non-empirical WOT-SRSH variant~\cite{wing_band_2021} achieves $\mathord{\sim}$0.1-0.2 eV gap accuracy for bulk materials~\cite{sagredo_electronic_2024,ohad_optical_2023,ohad_band_2022}.
The WOT-SRSH approach also transfers well to two-~\cite{camarasa-gomez_excitations_2024, florio_resolving_2025} and one-dimensional~\cite{ohad_nonempirical_2024} crystals, surfaces~\cite{sagredo_reliability_2025}, and point defects~\cite{ke_accurate_2025}.  
Its orbitals also provide excellent starting points for the $GW$ approximation~\cite{gant_optimally_2022, sagredo_electronic_2024}, BSE, and TDDFT approaches~\cite{ohad_optical_2023, sagredo_electronic_2024}.
%While the WOT-SRSH tuning ansatz has proven effective for a broad range of materials, its theoretical underpinnings and practical tunability are currently best understood in the context of non-magnetic materials.
%In particular, the calculation of the $N-1$ electron system when magnetic ordering is present is non-trivial, a subject of future work, and a current limitation of the method.

Broad application of the WOT-SRSH functional, including in high-throughput contexts, is currently limited in practice by a largely manual tuning protocol.
The conventional tuning procedure involves calculating a maximally-localized Wannier function~\cite{marzari_maximally_1997,souza_maximally_2001} representing the top of the valence manifold followed by a series of hybrid-DFT calculations for large supercells to tune and generate SRSH parameters.
During this tuning process, the selection of the short-range (SR) exact-exchange fraction $\alpha$ has relied on heuristic values, and depending on the value of the dielectric constant, can require ad hoc adjustment.
However, as has been done in the case of 2D materials~\cite{camarasa-gomez_excitations_2024, florio_resolving_2025}, $\alpha$ can be selected more deterministically albeit at higher computational cost using multiple phases of the same material.
In this work, we introduce an automated workflow that alleviates these bottlenecks and involves a well-defined and relatively cheap procedure for selecting the amount of SR exact-exchange that guarantees tunability.
Our implementation is currently limited to 3D bulk non-magnetic systems, where tunability is most robust, but it can straightforwardly be extended to 2D non-magnetic materials.
Additionally, seeing as the WOT-SRSH functional can only be tuned for gapped systems, it is not applicable to metallic systems.
Using this workflow, we find that the tuning objective function used by WOT-SRSH needs to be evaluated only three to five times for each material, as opposed to the typically more comprehensive sampling performed in preceding work.
By automating the tuning procedure and reducing its computational cost, we make WOT-SRSH more accessible and also scalable to large materials datasets, thereby enabling high-throughput applications.

\section{WOT-SRSH Functional}
The WOT-SRSH functional~\cite{wing_band_2021} is based on the screened range-separated hybrid (SRSH) functional formalism~\cite{refaely-abramson_gap_2013, refaely-abramson_solidstate_2015, kronik_excitedstate_2016}; it mixes a fraction of exact-exchange and semi-local exchange, as in a standard hybrid functional~\cite{becke_new_1993,perdew_rationale_1996}, but with different fractions in the short and long ranges. This is achieved by partitioning the exchange part of the Coulomb interaction using the identity
\begin{equation}
\label{eq:srsh_partition}
    \frac{1}{r}=\underbrace{\frac{\alpha+\beta\text{erf}(\gamma r)}{r}}_{\text{Fock}}
    +\underbrace{\frac{1-\left(\alpha+\beta\text{erf}(\gamma r)\right)}{r}}_{\text{(Semi-)Local}}.
\end{equation}
The left term is treated with a Fock-like exact-exchange operator, while the right term is treated by (semi-)local Kohn-Sham exchange~\cite{yanai_new_2004}.
The partition in Eq.~(\ref{eq:srsh_partition}) introduces three parameters: $\alpha$, $\beta$, and $\gamma$. $\alpha$ and $\alpha+\beta$ dictate the amount of exact-exchange in the SR and long-range (LR) limits, respectively, and $\gamma$ mediates the transition between these two limits~\cite{refaely-abramson_quasiparticle_2012, srebro_does_2012}.

The WOT-SRSH method owes its accuracy to the enforcement of two physical constraints in the SRSH functional via tuning $\alpha$, $\beta$, and $\gamma$.
First, it fixes the approximately-correct asymptotic long-range screening of the Coulomb potential in a material by setting~\cite{refaely-abramson_gap_2013}
\begin{equation}
\label{eq:dielectric_constrain}
    \alpha+\beta=1/\varepsilon_\infty,
\end{equation} where $\varepsilon_\infty$ is the directionally averaged static clamped-ion dielectric constant.
The second constraint is an ansatz~\cite{ma_using_2016} that generalizes~\cite{ohad_foundations_2025} the ionization potential theorem~\cite{perdew_densityfunctional_1982, levy_exact_1984, almbladh_exact_1985}, given by
\begin{equation}
    \label{eq:Delta_I_const}
    \braket{\phi_{w}|\hat{H}_{\text{SRSH}}(\alpha, \beta, \gamma)|\phi_{w}} = E_N(\alpha, \beta, \gamma) - \tilde{E}_{N-1}\left[\phi_w\right](\alpha, \beta, \gamma).
\end{equation}
In Eq.~(\ref{eq:Delta_I_const}), $\phi_{w}$ is a maximally-localized Wannier function, $\hat{H}_{\text{SRSH}}$ is the screened range-separated hybrid Hamiltonian, $E_N(\alpha, \beta, \gamma)$ is the total energy of the $N$ electron system, and $\tilde{E}_{N-1}(\alpha, \beta, \gamma)$ is the total energy from a constrained DFT calculation with $N-1$ electrons where the Wannier function $\phi_{w}$ is depopulated. We note here that although a Wannier function is used to satisfy the constraint in Eq.~\ref{eq:Delta_I_const}, the resultant functional using the determined optimal parameters has no need for said Wannier function.
In practice, $\tilde{E}_{N-1}(\alpha, \beta, \gamma)$ is computed in a supercell with a large Lagrange-like energy penalty coefficient $\lambda$ such that:
\begin{equation}
    \label{eq:E_N-1}   
    \tilde{E}_{N-1}(\alpha, \beta, \gamma)\left[\phi_w\right]=\min_{\{\psi_i\}}\left\{E_{N-1}\left(\alpha, \beta, \gamma,\{\psi_i\}\right)+\lambda\left(\sum_{i=1}^{N-1}|\braket{\psi_i|\phi_w}|^2-f_{\phi_w}\right)\right\}+E_{\text{img}}
\end{equation}
where $\psi_i$ is an eigenfunction of the constrained $N-1$ electron system, and $E_{\text{img}}$ is an image charge correction to offset spurious long-range interactions between removed electrons in periodic images of the supercell~\cite{leslie_energy_1985, makov_periodic_1995, komsa_finitesize_2012, rurali_theory_2009, lany_assessment_2008}.
Though an image charge correction is applied, the calculation is still carried out in supercells to reduce image charge effects in $\tilde{E}_{N-1}$.
Satisfying Eq.~(\ref{eq:Delta_I_const}) is achieved by fixing $\alpha$ and $\beta$ and tuning $\gamma$ to find the zero of the objective function $\Delta I(\alpha, \beta, \gamma)$, defined as
\begin{equation}
    \label{eq:delta_I}
    \Delta I(\alpha, \beta, \gamma)=\braket{\phi_{w}|\hat{H}_{\text{SRSH}}(\alpha, \beta, \gamma)|\phi_{w}} - E_N(\alpha, \beta, \gamma) + \tilde{E}_{N-1}[\phi_{w}](\alpha, \beta, \gamma).
\end{equation}
For further details on the WOT-SRSH functional, we refer the reader to Ref.~\cite{wing_band_2021}.

\begin{figure}[htb!]
    \centering
    \includegraphics[width=0.48\linewidth]{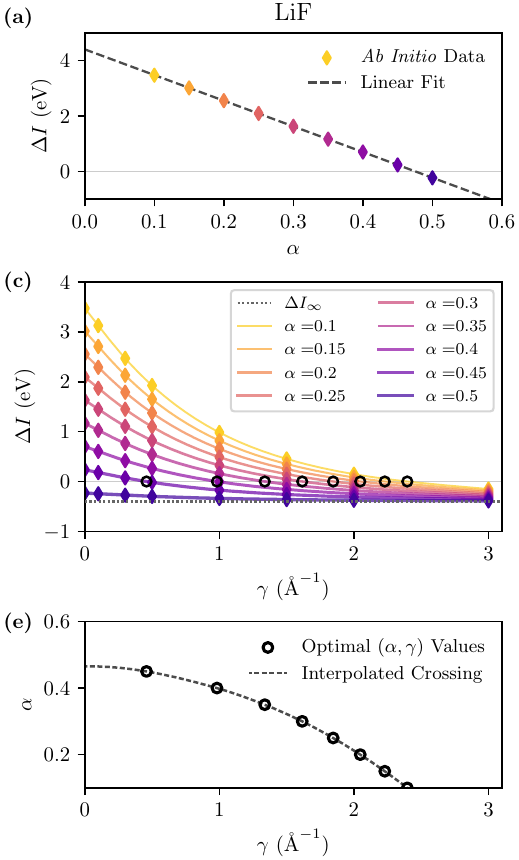}\includegraphics[width=0.48\linewidth]{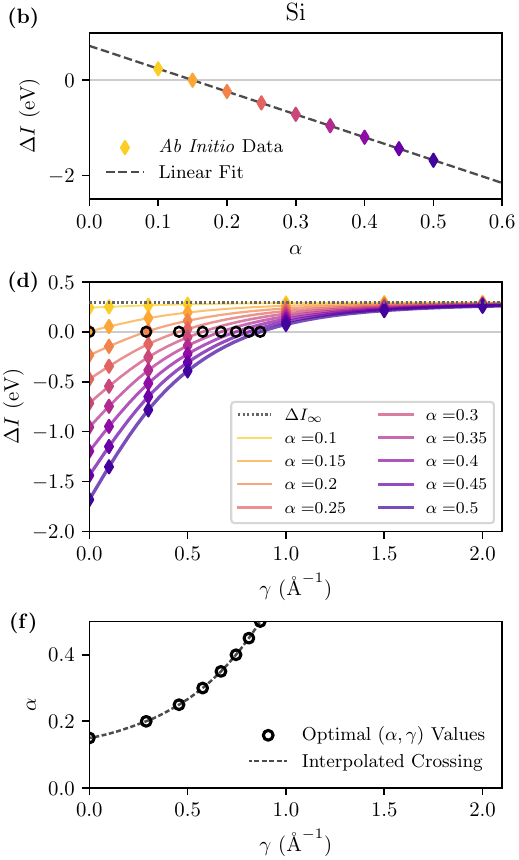}
    \caption{$\Delta I$ tuning landscape for LiF (left) and Si (right). Panels~\textbf{(a)}-\textbf{(b)} illustrate the near‐linear dependence of $\Delta I$ on $\alpha$ in the non-range-separated limit ($\beta=0$ and/or $\gamma=0$). Panels~\textbf{(c)}-\textbf{(f)} fix $\alpha + \beta = 1 / \varepsilon_{\infty}$, and panels~\textbf{(c)}-\textbf{(d)} show $\Delta I$ as a function of the range-separation parameter, $\gamma$, for fixed values of $\alpha$. All curves converging to the common asymptote $\Delta I_{LR}$ as $\gamma\to\infty$. Zero crossings, where $\Delta I=0$, are highlighted by black circles. Panels~\textbf{(e)}-\textbf{(f)} collect these zero‐crossings to show the one‐dimensional manifolds in $(\alpha,\gamma)$ space where the optimal tuning constraints are satisfied. The terminations of these curves at $\gamma=0$ indicate a critical value of $\alpha$, below or above which zero crossing will not occur.}\label{fig:alpha_ranges}
\end{figure}

\section{Parameter tuning}
Using LiF and Si as prototypical examples~\cite{refaely-abramson_solidstate_2015}, Fig.~\ref{fig:alpha_ranges} summarizes several key features of the $\Delta I$‑tuning landscape having established the long‑range screening constraint $\alpha+\beta=1/\varepsilon_\infty$.
Panels~(a) and~(b) show that in the non-range-separated limit ($\beta=0$ or $\gamma=0$ or $\gamma\to\infty$), $\Delta I$ depends nearly linearly on $\alpha$.  
Panels~(c) and~(d) leave this limit and show results obtained using SRSH functionals with $\beta\neq0$.
As $\gamma\to\infty$, all curves converge to the same asymptote $\Delta I_{LR}$, reflecting the reduction of each SRSH in this limit to a non-range-separated hybrid with a range-independent exact-exchange fraction, $\alpha+\beta=1/\varepsilon_\infty$.
Each colored curve in panels~(c) and~(d) tracks the evolution of $\Delta I$ with $\gamma$ for a specific $\alpha$, and the black circles identify the values of $\gamma$, for each $\alpha$, where $\Delta I$ crosses zero, satisfying Eq.~(\ref{eq:Delta_I_const}).  
Panels~(e) and~(f) collect these zero‑crossings, mapping out the one‑dimensional manifold of $(\alpha,\gamma)$ values where $\Delta I=0$ for LiF and Si, respectively.
End points at $\gamma=0$ reveal $\alpha$ ranges where $\Delta I$ never crosses zero (large $\alpha$ for LiF, small $\alpha$ for Si), for which optimal tuning is impossible; this is evident in the $\alpha=0.5$ and $\alpha=0.1$ curves of panels (c) and (d) respectively.

Altogether, Fig.~\ref{fig:alpha_ranges} suggests a new route for improving tuning efficiency, which is at the heart of the automated workflow suggested below.
First, since $\Delta I$ is essentially linear in $\alpha$ at $\gamma\rightarrow0$ but independent of $\alpha$ as $\gamma\to\infty$, the ranges of $\Delta I$ in these two limits can be mapped out with only a few calculations.
Second, there exists a continuous 1D curve of ``optimal'' $(\alpha,\gamma)$ pairs, because there are three free parameters in the SRSH formalism but only two constraints provided by Eqs.~(\ref{eq:dielectric_constrain}) and (\ref{eq:Delta_I_const}).
In prior work~\cite{wing_band_2021, ohad_band_2022, ohad_optical_2023, sagredo_electronic_2024, ohad_nonempirical_2024}, selecting an optimal $(\alpha,\gamma)$ pair along the curves shown in panels (e) and (f) has been done by manually choosing $\alpha$ so that $\gamma$ and $\beta$ can be tuned to satisfy Eqs.~(\ref{eq:dielectric_constrain}) and (\ref{eq:Delta_I_const}).
Alternatively, in the case of layered bulk materials that also have 2D phases, it is possible to select a single value of $\alpha$ by finding the crossing of the curves shown in panels (e) and (f) for the bulk and 2D phases~\cite{ramasubramaniam_transferable_2019, camarasa-gomez_transferable_2023}, a procedure that is non-empirical in the context of WOT-SRSH~\cite{camarasa-gomez_excitations_2024, florio_resolving_2025}.
However, neither of these approaches is robust for an automated search for optimal parameters as they either run the risk of choosing a value of $\alpha$ for which Eq.~(\ref{eq:Delta_I_const}) cannot be satisfied, or they can only be applied to a select class of materials and require an extensive sampling of $\Delta I$.
%Second, even if Eq.~\ref{eq:Delta_I_const} can be satisfied, the value of $\gamma$ for which $\Delta I=0$, can be quite small, making the SRSH functional nearly a global hybrid with exchange fraction $\alpha$ that does not practically enforce physically accurate asymptotic screening the LR limit.

Selecting an appropriate $\alpha$ value so that Eqs.~(\ref{eq:dielectric_constrain}) and (\ref{eq:Delta_I_const}) can be satisfied in practice motivates a novel third constraint for fixing the amount of SR exchange. In particular, $\alpha$ must be sufficiently large, or small, so that the SR limit of $\Delta I$ is opposite in sign to $\Delta I_{LR}$, ensuring there exists a value of $\gamma$ for which $\Delta I=0$.
In this work, we achieve this by choosing an optimal value of $\alpha$, $\alpha^{\text{opt}}$, to ensure that
\begin{equation}
\label{eq:alpha_constraint}
  \Delta I(\alpha=\alpha^{\text{opt}},\beta=\varepsilon_\infty^{-1}-\alpha^{\text{opt}},\gamma\to0)=-\Delta I(\alpha=\alpha^{\text{opt}},\beta=\varepsilon_\infty^{-1}-\alpha^{\text{opt}},\gamma\to\infty).
\end{equation}
In other words, we fix $\alpha$ so that the $\Delta I$ in the SR and LR limits are equal in magnitude but opposite in sign.
%this ensures a zero crossing in $\Delta I$ at a value of $\gamma$ which is not close to zero.

%Additionally, while optimal parameters can be determined for lower dimensional systems, this tuning process can be more involved and was therefore not the focus of this work.
%As such, we restrict the scope of this work to 3D non-magnetic bulk materials.

\section{Workflow Details}
\begin{figure}[htb!]
    \centering
    \includegraphics[width=0.9\linewidth]{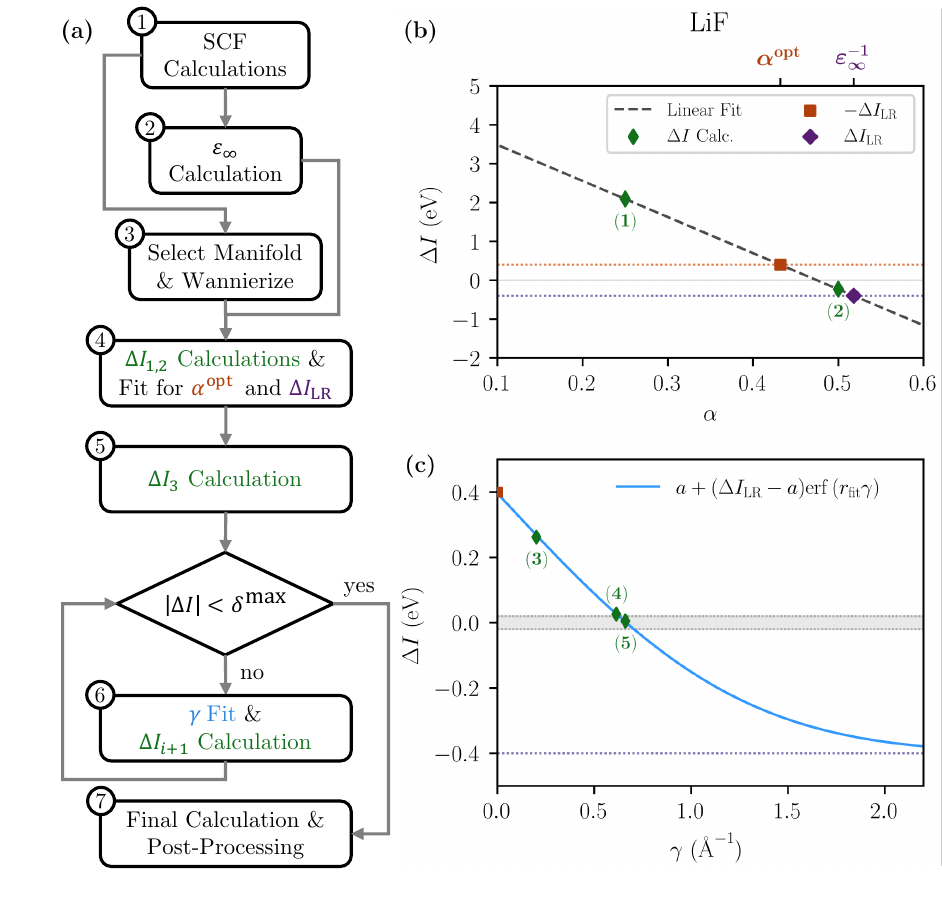}
    \caption{Automated WOT‑SRSH workflow.
    \textbf{(a)} Schematic of the seven step procedure: \circledcustom{0.0ex}{0.5pt}{1} perform an SCF unit cell calculation; \circledcustom{0.0ex}{0.5pt}{2} compute the clamped‐ion dielectric constant $\varepsilon_\infty$; \circledcustom{0.0ex}{0.5pt}{3} build a cubic supercell, generate maximally localized Wannier functions, and select the highest energy valence Wannier orbital, $\phi_w$; \circledcustom{0.0ex}{0.5pt}{4} determine the optimal exact‐exchange fraction $\alpha^{\rm opt}$ by two simultaneous non-range-separated hybrid $\Delta I$ calculations, at $\alpha=0.25$ and $0.50$, and a linear fit (green diamonds in panel b); \circledcustom{0.0ex}{0.5pt}{5} fix $\alpha=\alpha^{\rm opt}$ and $\beta=1/\varepsilon_\infty-\alpha^{\rm opt}$, sample $\Delta I$ at an initial $\gamma$ and test $|\Delta I|<\delta^{\max}$; \circledcustom{0.0ex}{0.5pt}{6} if un-converged, fit $\Delta I(\gamma)$ using Eq.~(\ref{eq:fit_func}), predict a new $\gamma$, and iterate; \circledcustom{0.0ex}{0.5pt}{7} once $|\Delta I|<\delta^{\max}$, finalize $(\alpha,\beta,\gamma)$ and proceed to band structure calculations.
    \textbf{(b)} Illustration of step four for LiF: green markers denote the two non-range-separated hybrid $\Delta I$ points, the dashed line is the linear fit, the orange square is $-\Delta I_{LR}$, and the purple diamond is the $\Delta I_{LR}$ asymptote; the crossing of the fit with $\Delta I=0$ yields $\alpha^{\rm opt}$.
    \textbf{(c)} Example of the $\gamma$‐tuning loop (steps 5-6): the blue curve is the erf fit through $\Delta I_1$ and $\Delta I_3$, and successive green markers show $\Delta I(\gamma)$ evaluations (steps 3-5) until convergence.}\label{fig:wot_workflow}
\end{figure}

Our workflow is shown in Fig.~\ref{fig:wot_workflow}.
Panel~(a) shows the overall workflow, and panels~(b), (c) show \textit{ab intio} data for steps four and six, respectively, for the case of LiF.

\textbf{First}, starting from a primitive unit‑cell structure, a larger supercell with a volume of at least $10^{3}$ $\text{\AA}^{3}$, constructed to be as close to cubic as possible, is generated using the Cubic Supercell Transformation class of the \texttt{pymatgen} package~\cite{ong_python_2013}. This is done to ensure that the $\tilde{E}_{N-1}$ calculation is done in a cell of sufficiently large volume to reduce image charge interaction effects. To account for the remaining image effects, we employ a Makov–Payne monopole image charge correction~\cite{makov_periodic_1995, wing_band_2021}.
Though not all supercells are perfectly cubic, we found the corrections obtained from a more generalized scheme~\cite{rurali_theory_2009} to be within 0.05~eV.
%The results of later DFT calculations in the supercell are used to Wannierize the selected manifold and extract the highest expectation energy Wannier function.

Concurrently, self-consistent DFT calculations for both the unit cell and the supercell are carried out using the PBE~\cite{perdew_generalized_1996} functional. If PBE is known to produce a spurious metallic ground state, the HSE06~\cite{krukau_influence_2006} functional is used instead in order to attempt to open up the band gap.
The results of this calculation are then used to determine manifold size of the isolated bands, for constructing the Wannier functions; the procedure for doing so is as follows.
Starting with the highest occupied eigenvalue at each $\bm{k}$-point, we record the lowest of these values.
Then, we consider the next-highest eigenvalue at every $\bm{k}$-point, identify the largest of these, and compare the value to the previously recorded value.
If the latter is lower in energy by more than $0.5$~eV, then the first band is considered to be isolated.
Otherwise, the same procedure as above is carried out for the second and third lowest energies at each $\bm{k}$-point and so on.
The search terminates when either an isolated manifold is found or when all bands have been looked through---in which case the isolated manifold is simply all the valence bands used in the calculation (e.g. Si, C).

\textbf{Second}, the static clamped‑ion dielectric tensor, $\varepsilon_\infty^{ij}$, is computed for the primitive unit cell using the converged density from step one.
We use the average of the trace of this quantity to set $\alpha+\beta$.
In the current workflow, the dielectric tensor is computed using the HSE06~\cite{krukau_influence_2006} functional, though PBE0~\cite{perdew_rationale_1996} or other hybrids could be employed in the event HSE06 fails to open a gap.
We note that it is possible to extend our workflow and use the tuned WOT-SRSH functional to calculate $\varepsilon_\infty$ in a self-consistent manner, as has been done in prior work~\cite{ohad_optical_2023}. This is not currently implemented in the workflow as the effects on final band gap MAE is only $\mathord{\sim}0.01$ eV.

\textbf{Third}, a set of maximally-localized Wannier functions (MLWF) is generated from a unitary transformation of the top of the valence band manifold in the supercell by using \texttt{Wannier90}~\cite{mostofi_updated_2014}. The Wannier function with the highest expectation energy value, $\phi_w$, is selected and used in subsequent steps.

Because only an isolated valence band manifold is used in this procedure, wannierization is performed without any disentanglement.
%By default, the first isolated manifold is selected and used for Wannierization. 
Although more sophisticated initial projections for this manifold, such as SCDM~\cite{vitale_automated_2020}, could be employed, we find that using Bloch projections is sufficient for all of the materials studied in this work.
In larger systems with more pronounced starting guess localization issues, SCDM or related methods could become necessary.
Section S1 of the Supplementary Material (SM) contains more details on the Wannierization procedure.

\textbf{Fourth}, we find $\alpha^{\text{opt}}$ by performing two simultaneous $\Delta I$ calculations ($\Delta I_1$ and $\Delta I_2$) in the non-range-separated hybrid limit with $\alpha=0.25$ and $\alpha=0.50$ (green diamonds in Fig.~\ref{fig:wot_workflow}~(b)).
Using the near-perfect linearity of $\Delta I$ in this non-range-separated hybrid limit (see Fig.~\ref{fig:alpha_ranges}~(a)-(b)), we fit a straight line through these two points.
From this fit, we extract the large‑$\gamma$ asymptote, $\Delta I_{LR}$, evaluated at $\alpha=1/\varepsilon_\infty$ as well as the optimal mixing fraction, $\alpha^{\text{opt}}$, satisfying $\Delta I(\alpha^{\rm opt})=-\Delta I_{LR}$.

We note that this approach works for determining $\alpha^{\text{opt}}$ even when $\Delta I_{LR}$ cannot be calculated explicitly, a scenario that can occur if $\varepsilon_\infty$ is large and the KS eigensystem is gapless at the semi-local level (e.g. InSb, InAs).
We also note that the nearly-perfect linear dependence of total energies and orbital eigenvalues on $\alpha$ can be rationalized from considering the limit that the DFT orbitals $\left\{\phi_{n\bm{k}}(\bm{r})\right\}$ remain fixed. In this limit, these quantities are \textit{exactly} linear in $\alpha$, but in practice, orbitals relax slightly with respect to changes in $\alpha$~\cite{miceli_nonempirical_2018}.
In practice, the workflow also possess robustness to variation in linearity in $\alpha$. Even if the extrapolated values of $\alpha^{\rm opt}$ and $\Delta I_{\rm LR}$ are slightly off, step 6 of the workflow continues to iterate until $\Delta I$ is below the desired convergence threshold.

\textbf{Fifth}, $\alpha$ and $\beta$ are fixed at $\alpha=\alpha^{\text{opt}}$ and $\beta=\varepsilon_\infty^{-1}-\alpha^{\text{opt}}$, making $\Delta I$ a function of $\gamma$ only. Using an initial guess of $\gamma=0.2$\,\text{\AA}$^{-1}$, $\Delta I$ is sampled a third time, giving $\Delta I_3$.
If $\lvert\Delta I(\gamma=0.2\,\text{\AA}^{-1})\rvert$ is below the numerical tolerance, $\delta^{\text{max}}=0.02$~eV, the tuning procedure is complete. 

\textbf{Sixth}, if $\lvert\Delta I(\gamma=0.2\,\text{\AA}^{-1})\rvert\geq\delta^{\text{max}}$ convergence is not yet achieved; tuning continues.
The values of $\Delta I_{LR}$, $-\Delta I_{LR}$, and $\Delta I_3$, are used fit to the function
\begin{equation}
\label{eq:fit_func}
  \Delta I^{\text{fit}}(r_{\text{fit}}\,\gamma)=a+\left(\Delta I_{LR}-a\right)\,\mathrm{erf}\left(r_{\text{fit}}\,\gamma\right),  
\end{equation}
where $a$, and $r_{\text{fit}}$ are free parameters.
The general form of this function is meant to approximate the total amount of exact-exchange present in the SRSH functional, a quantity that tends to correspond linearly to the value of $\Delta I$. The exact LR limit of $\Delta I$ is enforced by explicitly including the value of $\Delta I_{LR}$ in the fit ansatz.
Given the values of $r_{\text{fit}}$ and $a$ from the fit, the zero of $\Delta I^{\text{fit}}$ provides a new optimal $\gamma$ prediction, namely $\gamma=\mathrm{erf}^{-1}\big(\frac{a}{a-\Delta I_{LR}}\big)r_{\text{fit}}^{-1}$.
This value is then used in a new calculation of $\Delta I$, and a fit‑and‑update loop is repeated until the $\lvert\Delta I(\gamma)\rvert<\delta^{\text{max}}$ criterion is met.
We also remark that other means of finding an optimal $\gamma$ could be employed, but the fit given by Eq.~(\ref{eq:fit_func}) has proven quite effective for all of the systems studied in this work.

\textbf{Seventh}, once the parameters $(\alpha,\beta,\gamma)$ are tuned to satisfy Eqs.~(\ref{eq:dielectric_constrain}), (\ref{eq:Delta_I_const}), and (\ref{eq:alpha_constraint}), the resulting WOT-SRSH functional can be used to calculate the band structure, or any other property of interest, at the same computational cost as other hybrid functionals with LR exact-exchange.

The workflow presented here is implemented via the \texttt{Jobflow} library~\cite{rosen_jobflow_2024} and uses an in-house modification of \texttt{VASP}~\cite{kresse_initio_1993,kresse_initio_1994,kresse_normconserving_1994,kresse_efficiency_1996,kresse_efficient_1996,kresse_ultrasoft_1999}.
%that contains the implementation of the SRSH functional as its \textit{ab initio} code and a constrained minimization procedure for Wannier-optimal tuning.
These modifications are, broadly speaking, twofold.
First, we have introduced the generalized RSH functional, allowing for the free variation of $\alpha$, $\beta$, and $\gamma$.
And second, we have implemented the constrained minimization procedure described in Eq.~\ref{eq:E_N-1} using an approach similar to DFT+U.
%Currently, this patch is not yet available publicly, but it can be obtained after request.
%(or available publicly), {\color{BrickRed}but it will be directly available within the official \texttt{VASP} soure code in the near future}.
These choices ensure a high level of computational efficiency.
First, the use of \texttt{Jobflow} is implemented so that independent tasks, such as the calculation of the dielectric constant and the Wannierization (steps 2 and 3), as well as the sampling of $\Delta I$ which requires the calculation of the ground state energies for both the $N$ and $N-1$ electron systems, can be run in parallel.
%However, the tuning calculations for sampling of the $i$th step of the $\Delta I_i$ function are run in sequence as each one depends on the set of prior calculations.
Second, the use of \texttt{VASP} to calculate WOT-SRSH parameters benefits from the computational efficiency of projector augmented wave method~\cite{kresse_ultrasoft_1999}, for providing an effective description of the effect of core electrons.
However, this workflow can be readily implemented in other \textit{ab initio} codes.

The workflow can be generated in a few lines of code and its input parameters are provided in a YAML file (see section S2 of the SM). The workflow makes use of the VASP input sets provided by the \texttt{pymatgen} package, but we note that the user can also overwrite these input parameters by editing the dedicated input YAML file per calculation type or per code. 

%More practical info and some tutorials about how to use the workflow are provided at \url{https://github.com/fraricci/wotsrsh-workflow/tree/vasp_jobflow}.

\section{Results}
We benchmark our workflow on 23 semiconductors and insulators---including a number of complex metal oxides---originally used in Refs.~\cite{wing_band_2021} and~\cite{ohad_optical_2023} to evaluate the performance of the WOT-SRSH functional.
The main results for the workflow tuning comparison are reported in Tables~\ref{tab:known_sc} and~\ref{tab:metal_oxides}; they contain the fundamental band gap, parameters for the SRSH functional, and the number of times $\Delta I$ is sampled, $N_{\Delta I}$, for each material.
For the sake of comparison, we use the same dielectric constants computed in Refs.~\cite{wing_band_2021, ohad_optical_2023}, but a comparison of the workflow's computed values of this as well as the resulting gaps is given in section S3 of the SM.
Additional results, including computational cost summaries, plots of the $\Delta I$ vs $\gamma$ tuning curves, and the WOT-SRSH band structures for each material are reported in sections S5, S6, and S7 of the SM, respectively.

The crystal structures of the materials used in our calculations are obtained from experiment and provided in section S8 of the SM. Alongside these data, the sizes of the supercells used in the tuning of $\Delta I$ for each material are also reported in S8.
More information on the Wannier functions used in tuning can be found in section S1 of the SM. There, we report the number of unit cell bands used to construct the Wannier functions, as well as the spread of the selected top valence Wannier functions used in the tuning.

\subsection{Tuned Parameters and Computed Band Gaps}

\begin{table}[htb!]
\centering
\caption{\label{tab:known_sc} Fundamental band gaps for semiconductors and insulators considered in Ref.~\cite{wing_band_2021} with the corresponding tuned $\alpha$ and $\gamma$ parameters. The computed values found in this work (WFlow) are compared with those from Ref.~\cite{wing_band_2021}. Experimental values (Exp.) for the fundamental band gaps are taken from various sources, see section S4 in the SM; the dielectric constant $\epsilon_\infty$ values are taken from Ref.~\cite{wing_band_2021}. $N_{\Delta I}$ is the number of computed $\Delta I$ needed to find the tuned $\alpha$ and $\gamma$ parameters. The bottom row displays both the band gap mean absolute error (MAE) and the maximum error (ME) compared to the experimentally obtained fundamental band gap. The difference in the MAE between the manual tuning as in Ref.~\cite{wing_band_2021} and the workflow presented here is only 0.03~eV.}
\begin{threeparttable}
\begin{tabular}{l|ccc|cc|cc|c|c}
\hline
& \multicolumn{3}{c|}{Fund. Band gap (eV)} & \multicolumn{2}{c|}{$\alpha$} & \multicolumn{2}{c|}{$\gamma$ (1/\AA)} & $\varepsilon_\infty$ & {\centering$N_{\Delta I}$}\\
& \cite{wing_band_2021} & WFlow & Exp. & WFlow & \cite{wing_band_2021} & WFlow & \cite{wing_band_2021}  & \cite{wing_band_2021} & WFlow\\ \hline
InSb & 0.32\tnote{\color{RoyalBlue}\dag} & 0.30\tnote{\color{RoyalBlue}\dag} & 0.20 & 0.22 & 0.25 & 0.24 & 0.32  & 13.24 & 4 \\
InAs & 0.42\tnote{\color{RoyalBlue}\dag} & 0.46\tnote{\color{RoyalBlue}\dag} & 0.37 & 0.23 & 0.25 & 0.26 & 0.30  & 11.40 & 4 \\
Ge & 0.69\tnote{\color{RoyalBlue}\dag} & 0.77\tnote{\color{RoyalBlue}\dag} & 0.71 & 0.27 & 0.25 & 0.33 & 0.36  & 14.79 & 5 \\
GaSb & 0.69\tnote{\color{RoyalBlue}\dag} & 0.73\tnote{\color{RoyalBlue}\dag} & 0.78 & 0.23 & 0.25 & 0.24 & 0.36  & 13.04 & 4 \\
Si & 1.14 & 1.19 & 1.19 & 0.21 & 0.25 & 0.33 & 0.45  & 11.25 & 4 \\
InP & 1.56\tnote{\color{RoyalBlue}\dag} & 1.57\tnote{\color{RoyalBlue}\dag} & 1.39 & 0.23 & 0.25 & 0.27 & 0.43  & 8.87 & 4 \\
GaAs & 1.41\tnote{\color{RoyalBlue}\dag}  & 1.45\tnote{\color{RoyalBlue}\dag}  & 1.48 & 0.25 & 0.25 & 0.27 & 0.28  & 10.52 & 4 \\
AlSb & 1.71\tnote{\color{RoyalBlue}\dag}  & 1.79\tnote{\color{RoyalBlue}\dag}  & 1.65 & 0.24 & 0.25 & 0.24 & 0.26  & 9.82 & 4 \\
AlAs & 2.25\tnote{\color{RoyalBlue}\dag}  & 2.30\tnote{\color{RoyalBlue}\dag}  & 2.19 & 0.28 & 0.25 & 0.27 & 0.19  & 8.19 & 4 \\
GaP & 2.39\tnote{\color{RoyalBlue}\dag} & 2.44\tnote{\color{RoyalBlue}\dag} & 2.36 & 0.24 & 0.25 & 0.28 & 0.40  & 8.89 & 4 \\
AlP & 2.52 & 2.63 & 2.47 & 0.27 & 0.25 & 0.28 & 0.30  & 7.29 & 4 \\
GaN & 3.76 & 3.68 & 3.61 & 0.26 & 0.30 & 0.38 & 0.45  & 5.03 & 4 \\
C    & 5.76 & 5.78 & 5.84 & 0.31 & 0.30 & 0.50 & 0.43  & 5.55 & 5 \\
AlN & 6.56 & 6.50 & 6.53 & 0.31 & 0.35 & 0.39 & 0.49  & 4.12 & 4 \\
MgO & 8.16 & 8.09 & 8.30 & 0.34 & 0.25 & 0.20 & 2.83  & 2.90 & 3 \\
LiF & 15.34 & 15.04 & 15.35 & 0.43 & 0.25 & 0.66 & 2.04  & 1.93 & 5 \\ \hline
MAE & 0.07 & 0.11 &  &  &  &  &  &  &  \\
ME & 0.17 & 0.31 &  &  &  &  &  &  &  
\end{tabular}
\begin{tablenotes}
\footnotesize
\item[\color{RoyalBlue}\dag] Spin-orbit-coupling corrections were applied to these computed band gaps.
\end{tablenotes}
\end{threeparttable}
\end{table}

In Table~\ref{tab:known_sc}, we report the computed fundamental band gaps, as well as the tuned WOT-SRSH parameters for the semiconductors and insulator in Ref.~\cite{wing_band_2021}.
We compare our results to those reported in Ref.~\cite{wing_band_2021}, as well as to experiment in the case of the fundamental band gap.
The experimental reference gaps shown in Table~\ref{tab:known_sc} are measured room temperature fundamental band gaps that have been corrected to remove vibrational effects (i.e., zero-point and finite temperature gap renormalization of the fundamental band gap).
These corrected reference gap values comprise a rigorous benchmark set for our computed fundamental bands gaps.

Overall, we find excellent agreement between our automatically generated WOT-SRSH functionals and prior results.
Both approaches have an MAE relative to experiment of $0.1$ eV or less, and the automated generation of WOT-SRSH parameters only increases the MAE by $0.04$ eV.
The maximum error (ME) is slightly larger at $0.17$ and $0.31$ eV for the data of Ref.~\cite{wing_band_2021} and the workflow respectively.
In a few cases, namely for MgO and LiF, the differences between Ref.~\cite{wing_band_2021} and the workflow and/or between experiment and the workflow exceed $0.2$ eV.
This can be partly explained by the fact that the application of WOT-SRSH in Ref.~\cite{wing_band_2021} used a fixed value of $\alpha=0.25$ for all the materials, unless Eq.~(\ref{eq:Delta_I_const}) could not be satisfied---in which case it was increased to either $0.30$ or $0.35$. In contrast, the automated method employed here selects a generally different value of $\alpha$ for each system. As such, we observe that the $\alpha$ and $\gamma$ values found by
our approach are close but not identical to those reported previously. In particular, MgO and LiF exhibit a sizable $\alpha$ discrepancy; our workflow leads to $0.34$ and $0.43$, respectively, while both were $0.25$ in Ref.~\cite{wing_band_2021}. The larger value of $\alpha$ used in the workflow also results in much smaller values of $\gamma$ being needed to satisfy Eq.~\ref{eq:Delta_I_const}. Specifically, we find $\gamma=0.20\,\text{\AA}^{-1},\;0.66\,\text{\AA}^{-1}$ for MgO and LiF respectively, while in Ref.~\cite{wing_band_2021} $\gamma=2.83\,\text{\AA}^{-1},\;2.04\,\text{\AA}^{-1}$ for the same systems.
This discrepancy in optimal parameters is a manifestation of the fact that there exists a 1D subspace of the three SRSH parameters which can satisfy the constraints of Eqs.~(\ref{eq:dielectric_constrain}) and (\ref{eq:Delta_I_const}). As has been previously reported, the variation of the fundamental band gap within this subspace can be as large as $0.32$ eV in the case of AlN, though it is usually $\mathord{\sim}0.1$ eV or less~\cite{wing_band_2021, gant_optimally_2022}.
%As it is shown by our results, the band gap computed by our automated workflow matches the value reported in Ref. for 11 cases.
Other sources of error include different convergence criteria (here $\delta_{\max}=20$ meV), as well as challenges in accurately computing the effects of zero-point lattice motion on band gap renormalization.
Finally, we note that in the cases of the large-gap insulators LiF and MgO, though the absolute errors with respect to experiment are larger, their relative errors are only 2.3\% and 3.1\%, respectively.

In Table~\ref{tab:metal_oxides}, we report the computed fundamental band gap, as well as the tuned WOT-SRSH parameters, for the metal oxides considered in Ref.~\cite{ohad_optical_2023}.
\begin{table}[phtb!]
\centering
\caption{Fundamental band gaps for the metal-oxides considered in Ref.~\cite{ohad_optical_2023} with the corresponding tuned $\alpha$ and $\gamma$ parameters. The computed values found in this work (WFlow) are compared with those from Ref.~\cite{ohad_optical_2023}. Experimental values (Exp.) for the fundamental band gaps are taken from various sources, see S4 in the SM; the dielectric constant $\epsilon_\infty$ values are taken from Ref.~\cite{ohad_optical_2023}. $N_{\Delta I}$ is the number of computed $\Delta I$ needed to find the tuned $\alpha$ and $\gamma$ parameters. The bottom row displays the band gap mean absolute error (MAE) and the maximum error (ME) compared to the experimentally obtained fundamental band gap. The difference in the MAE between the manual manual tuning as in Ref.~\cite{ohad_optical_2023} and the workflow presented here is only 0.02~eV.}\label{tab:metal_oxides}
\begin{tabular}{l|ccc|cc|cc|c|c}
\hline
 & \multicolumn{3}{c|}{Fund. Band gap (eV)} & \multicolumn{2}{c|}{\(\alpha\)} & \multicolumn{2}{c|}{\(\gamma\) (1/\AA)} & $\varepsilon_\infty$ & $N_{\Delta I}$ \\
 & \cite{ohad_optical_2023} & WFlow & Exp. & WFlow & \cite{ohad_optical_2023} & WFlow & \cite{ohad_optical_2023}  & \cite{ohad_optical_2023} & WFlow  \\ \hline
Cu\textsubscript{2}O & 2.02 & 2.04 & 2.21 & 0.25 & 0.25 & 0.89 & 0.95   & 6.51 & 5 \\
BaSnO\textsubscript{3} & 3.46 & 3.48 & 3.34 & 0.26 & 0.30 & 0.20 & 1.40   & 3.92 & 3 \\
TiO\textsubscript{2} & 3.48 & 3.40 & 3.37 & 0.21 & 0.25 & 0.47 & 0.85   & 6.25 & 4 \\
BiVO\textsubscript{4} & 3.50 & 3.49 & 3.42 & 0.19 & 0.25 & 0.55 & 2.00   & 5.92 & 4 \\
ZnO & 3.53 & 3.49 & 3.53 & 0.29 & 0.30 & 0.20 & 1.30   & 3.57 & 3 \\
CaO & 6.61 & 6.54 & 6.74 & 0.29 & 0.25 & 0.52 & 1.70   & 3.25 & 4 \\
Al\textsubscript{2}O\textsubscript{3} & 9.80 & 9.92 & 9.77 & 0.37 & 0.40 & 0.47 & 1.40   & 2.94 & 4 \\ \hline
MAE & 0.09 & 0.11 &  &  &  &  &  &   &  \\
ME & 0.19 & 0.20 &  &  &  &  & &   & 
\end{tabular}
\end{table}
\noindent
We again find excellent agreement between the automatically generated WOT-SRSH parameters and prior benchmark results.
Both sets of parameters have a mean absolute error (MAE) relative to the experimental reference values of $0.1$ eV, and our generated WOT-SRSH parameters only increases the MAE by $0.02$ eV.
Overall, the workflow results in $\alpha$ values that are relatively close to those reported in Ref.\ \cite{ohad_optical_2023}, but tend to lead to reduced $\gamma$ values with respect to those in the same reference.
It is worth emphasizing how accurate the computed band gaps of these systems are, given how relatively inexpensive the tuning process is for them.
Many of the materials in Table~\ref{tab:metal_oxides}, such as ZnO, Cu\textsubscript{2}O, or BiVO$_4$ are well-known as being particularly difficult and expensive to describe accurately with other hybrid functionals or with the $GW$ approximation~\cite{usuda_allelectron_2002, rinke_combining_2005, vanschilfgaarde_quasiparticle_2006, vanschilfgaarde_adequacy_2006, fuchs_quasiparticle_2007, shishkin_selfconsistent_2007, schleife_bandstructure_2009, shih_quasiparticle_2010, friedrich_band_2011, stankovski_g0w0_2011, friedrich_hybrid_2012, samsonidze_insights_2014, wiktor_comprehensive_2017, golze_gw_2019, rangel_reproducibility_2020}.
We also note that compared to the results of the previous Table, where experimental band gap reference data were primarily obtained using techniques where excitonic effects are accurately accounted for in the measurement of the fundamental electronic band gap, the experimental data in Table~\ref{tab:metal_oxides} come primarily from optical absorption data and are subject to more uncertainty due to the need to fit for and subtract away excitonic effects in determining the reference gaps.

We also analyze the effects of calculating the static dielectric constant in the workflow itself using HSE06 as shown in Fig.~\ref{fig:wot_workflow}~(a).
The full results of these data are reported in Tables S5 and S6 of section S3 in the SM.
Overall, we calculate that using the single-shot HSE06 static dielectric constants only increases the MAE by 0.03 eV for the materials analyzed in Table~\ref{tab:known_sc} and 0.02 eV for the materials in Table~\ref{tab:metal_oxides}.

The computational cost of the tuning procedure, as well as final SCF and band structure calculations, are reported in Fig.~S1 in section S5 of the SM.
For convenience, these calculations are carried out for all the materials studied in the work on 4 nodes with 128 CPU cores in a few hours at most---see section S5 of the SM for more details.
The most expensive steps among all of the calculations used to determine the WOT-SRSH functional are those needed to obtain the dielectric tensor followed by the sampling of $\Delta I$, a process that requires $\mathord{\sim}$6--8 hybrid functional calculations in a supercell.
The main factor that influences the cost of these steps is the number of electrons in the system and the number of $\bm{k}$-points needed.
For systems that require spin-orbit coupling (SOC) corrections to the fundamental band gap, the final DFT and band structure calculations are more expensive due to the larger number of bands and denser $\bm{k}$-mesh required.
For systems without SOC, these calculations actually required the least resources.
Overall, even if this computational setting might be suboptimal, we find that the average time to carry out the entire workflow is $\mathord{\sim}$1.2 hours on 4 nodes and 128 CPU cores.

\subsection{Effects of Valence Manifold Selection Size}
\label{subsec:wannier_effects}
In order to assess the impact of Wannier function selection, we analyze the effects of including (when possible) a second larger isolated manifold in the Wannierization procedure.
This second manifold is determined by adding another band to the manifold selected in step one and then carrying out the same aforementioned selection procedure.
\begin{table}[htbp!]
\centering
\caption{\label{tab:known_sc_mani_comp}Tuned $\alpha$ and $\gamma$ parameters and the computed fundamental band gap for the semiconductors and insulators considered in Ref.~\cite{wing_band_2021}. Reference gap values are a combination of experimental band gap values taken from literature and ZPR corrections, as discussed in Ref.~\cite{wing_band_2021} and S4. The results of using the first vs first two isolated manifolds are reported here; both provide similar results overall.}
\begin{threeparttable}
\begin{tabular}{l|ccc|cc|cc}
\hline
 & \multicolumn{3}{c|}{Band gap (eV)} & \multicolumn{2}{c|}{$\alpha$} & \multicolumn{2}{c}{$\gamma$ (1/\AA)} \\ \hline
$N_{\text{manifold}}$ & 1 & 2 & Ref. & 1 & 2 & 1 & 2 \\ \hline
InSb & 0.30\tnote{\color{RoyalBlue}\dag} & 0.42\tnote{\color{RoyalBlue}\dag} & 0.20 & 0.22 & 0.27 & 0.24 & 0.29 \\
InAs & 0.46\tnote{\color{RoyalBlue}\dag} & 0.52\tnote{\color{RoyalBlue}\dag} & 0.37 & 0.23 & 0.28 & 0.26 & 0.31 \\
Ge & 0.77\tnote{\color{RoyalBlue}\dag} & 0.77\tnote{\color{RoyalBlue}\dag} & 0.71 & 0.27 & 0.28 & 0.33 & 0.34 \\
GaSb & 0.73\tnote{\color{RoyalBlue}\dag} & 0.86\tnote{\color{RoyalBlue}\dag} & 0.78 & 0.23 & 0.28 & 0.24 & 0.30 \\
GaAs & 1.45\tnote{\color{RoyalBlue}\dag} & 1.54\tnote{\color{RoyalBlue}\dag} & 1.48 & 0.25 & 0.29 & 0.27 & 0.32 \\
InP & 1.57\tnote{\color{RoyalBlue}\dag} & 1.56\tnote{\color{RoyalBlue}\dag} & 1.39 & 0.23 & 0.24 & 0.27 & 0.32 \\
AlSb & 1.79\tnote{\color{RoyalBlue}\dag} & 1.86\tnote{\color{RoyalBlue}\dag} & 1.65 & 0.24 & 0.29 & 0.24 & 0.29 \\
AlAs & 2.30\tnote{\color{RoyalBlue}\dag} & 2.31\tnote{\color{RoyalBlue}\dag} & 2.19 & 0.28 & 0.31 & 0.27 & 0.33 \\
GaP & 2.44\tnote{\color{RoyalBlue}\dag} & 2.42\tnote{\color{RoyalBlue}\dag} & 2.36 & 0.24 & 0.26 & 0.28 & 0.33 \\
AlP & 2.63 & 2.59 & 2.47 & 0.27 & 0.27 & 0.28 & 0.33 \\
GaN & 3.68 & 3.71 & 3.61 & 0.26 & 0.27 & 0.38 & 0.39 \\
AlN & 6.50 & 6.55 & 6.53 & 0.31 & 0.34 & 0.39 & 0.48 \\
MgO & 8.09 & 8.22 & 8.30 & 0.34 & 0.37 & 0.20 & 0.51 \\
LiF & 15.04 & 15.23 & 15.35 & 0.43 & 0.47 & 0.66 & 0.68 \\ \hline
MAE\tnote{\color{RoyalBlue}\ddag} & 0.12 & 0.11 &  &  &  &  &  \\
ME\tnote{\color{RoyalBlue}\ddag} & 0.31 & 0.22 &  &  &  &  &
\end{tabular}
\begin{tablenotes}
\footnotesize
\item[\color{RoyalBlue}\dag] As in Ref.~\cite{wing_band_2021}, spin-orbit-coupling corrections were applied to these computed band gaps.
\item[\color{RoyalBlue}\ddag] Values of C and Si are excluded from this analysis
\end{tablenotes}
\end{threeparttable}
\end{table}
\begin{table}[htbp!]
\centering
\caption{Tuned $\alpha$ and $\gamma$ parameters and the computed fundamental band gap for the metal-oxides considered in Ref.~\cite{ohad_optical_2023}. Reference gap values are a combination of experimental band gap values taken from literature and ZPR corrections, as discussed in S4. The results of using the first vs first two isolated manifolds are reported here; using the first manifold only has a mean absolute error (vs. experiment) that is 0.1 eV lower than using two.}
\label{tab:metal_oxides_mani_comp}
\begin{tabular}{l|ccc|cc|cc}
\hline
 & \multicolumn{3}{c|}{Band gap (eV)} & \multicolumn{2}{c|}{\(\alpha\)} & \multicolumn{2}{c}{\(\gamma\) (1/\AA)} \\ \hline
$N_{\text{manifold}}$ & 1 & 2 & Ref. & 1 & 2 & 1 & 2 \\ \hline
Cu\textsubscript{2}O & 2.04 & 2.04 & 2.21 & 0.25 & 0.25 & 0.89 & 0.89 \\
BaSnO\textsubscript{3} & 3.48 & 3.50 & 3.34 & 0.26 & 0.27 & 0.2 & 0.46 \\
TiO\textsubscript{2} & 3.40 & 3.48 & 3.37 & 0.21 & 0.24 & 0.47 & 0.53 \\
BiVO\textsubscript{4} & 3.49 & 3.49 & 3.42 & 0.19 & 0.19 & 0.55 & 0.60 \\
ZnO & 3.49 & 4.04 & 3.53 & 0.29 & 0.53 & 0.2 & 0.96 \\
CaO & 6.54 & 6.68 & 6.74 & 0.29 & 0.32 & 0.52 & 0.43 \\
Al\textsubscript{2}O\textsubscript{3} & 9.92 & 10.04 & 9.77 & 0.37 & 0.40 & 0.4717 & 0.55 \\ \hline
MAE & 0.11 & 0.20 &  &  &  &  &  \\
ME & 0.20 & 0.51 &  &  &  &  & 
\end{tabular}
\end{table}
The size of these manifolds, as well as the spreads of the selected ``top valence'' Wannier function for both cases, are reported in Table S1 in the SM. In said table we also report the fraction of contribution from the second-lowest manifold to the selected Wannier function.
Overall, we find that including a second lower manifold in the Wannierization yields more localized Wannier functions.
As summarized in Tables~\ref{tab:known_sc_mani_comp}, \ref{tab:metal_oxides_mani_comp}, the accuracy of band gaps computed with WOT-SRSH functionals tuned using these larger-manifold Wannnier functions is reduced.
The MAE using this approach for all materials considered increases by 0.03~eV, likely because the selected Wannier function hybridizes with deeper valence states which are not representative of the valence band maximum.
These findings indicate that the workflow is relatively insensitive to the choice of manifold, but using the smallest isolated manifold tends to provide higher-accuracy results.

%\begin{table}[htbp!]
%    \centering
%    \caption{Data on the error in computed band gaps for the number of isolated valence band manifolds, $N_{\text{manifold}}$, for all materials except Si and C. These systems were excluded because there was no second manifold to compare with due all 4 occupied valence bands being entabgled for both. Overall, using one versus two isolated manifold gives very similar results with the first highest manifold leading to more accurate results.}
%    \label{tab:manifold_errors}
%    \begin{tabular}{l|cc}
%    \hline
%    $N_{\text{manifold}}$ & 1 & 2 \\ \hline
%    MAE (eV) &  0.11 & 0.14 \\
%    ME (eV) &  0.31 & 0.51 \\ \hline
%    \end{tabular}
%\end{table}

\section{Conclusion}
We present a fully automated workflow for the non-empirical tuning of the Wannier optimally-tuned screened range-separated hybrid functional that removes previously existing ambiguities in optimal parameter selection and minimizes the cost of the tuning process.
By introducing a new and efficient sampling of the space of parameters that describe the SRSH functional and enforcing the ionization potential ansatz---often requiring only three to five evaluations---the protocol outlined here significantly reduces the computational (and human) overhead relative to previous manual approaches.
We find that our automated procedure successfully determines optimal values of the SRSH functional that yield computed band gaps in close agreement with experiment and prior benchmark studies, for a diverse set of semiconductors and insulators including complex metal oxides.
Compared to these benchmarks, the mean absolute error relative to experimental data is only slightly increased by no more than $0.04$ eV, and overall agreement with experimental band gap data remains within $\mathord{\sim}0.1$ eV.
In summary, the automated workflow developed here offers a robust and computationally efficient method for optimally tuning SRSH hybrid functionals to obtain experimentally accurate band gaps.
This advancement lays the groundwork for future developments, such as the integration of machine learning techniques to further accelerate parameter tuning and the possibility of high-throughput applications of the WOT-SRSH functional to a wide variety of materials.

\section*{Acknowledgments}
We thank Sijia Ke, Francisca Sagredo, and Jaydon Lee for insightful discussions and initial testing of the workflow. We also thank 
Dahvyd Wing and Jonah Haber for their help in developing earlier versions of the functional tuning implementation. This work was primarily supported by the Office of Science, Basic Energy Sciences, Materials Sciences and Engineering Division, under Contract No. DE-AC02-05CH11231 (Materials Project program KC23MP). Work at the Weizmann Institute was supported by the Israel Science Foundation. S.E.G. was partially supported by the Kavli Energy NanoScience Institute. Computational resources were provided by the Texas Advanced Computing Center (TACC) using the Frontera system through the allocation DMR23008 as well as by the National Energy Research Scientific Computing Center (NERSC), DOE Office of Science User Facilities supported by the Office of Science of the US Department of Energy under Contract DE-AC02-05CH11231. F.R. acknowledges support from the BEWARE scheme of the Wallonia-Brussels Federation for funding under the European Commission's Marie Curie-Skłodowska Action (COFUND 847587). A.R. acknowledges support from the National Science Foundation (NSF-BSF 2150562). L.K. acknowledges additional support from the Aryeh and Mintzi Katzman Professorial Chair and from the Helen and Martin Kimmel Award for Innovative Investigation.

\nocite{vurgaftman_band_2001, madelung_semiconductors_2004, kondo_structural_2008, mizoguchi_strong_2004, sleight_crystal_1979, werner_highpressure_1982, recker_directional_1988, sugiyama_crystal_1991, garcia-martinez_microstructural_1993, cardona_isotope_2005, miglio_predominance_2020, strossner_dependence_1986, clark_intrinsic_1997, piacentini_thermoreflectance_1976, nery_quasiparticles_2018, whited_exciton_1969, french_interband_1994, park_applicability_2022, aggoune_consistent_2022, lushchik_multiplication_1994, komornicki_structural_2004, wu_firstprinciples_2018, cooper_indirect_2015, wiktor_comprehensive_2017, wang_temperature_2003}

\bibliographystyle{elsarticle-num}
\bibliography{reference}

\begin{thebibliography}{100}
\expandafter\ifx\csname url\endcsname\relax
  \def\url#1{\texttt{#1}}\fi
\expandafter\ifx\csname urlprefix\endcsname\relax\def\urlprefix{URL }\fi
\expandafter\ifx\csname href\endcsname\relax
  \def\href#1#2{#2} \def\path#1{#1}\fi

\bibitem{hybertsen_firstprinciples_1985}
M.~S. Hybertsen, S.~G. Louie, First-{{Principles Theory}} of {{Quasiparticles}}: {{Calculation}} of {{Band Gaps}} in {{Semiconductors}} and {{Insulators}}, Phys. Rev. Lett. 55~(13) (1985) 1418--1421.
\newblock \href {https://doi.org/10.1103/PhysRevLett.55.1418} {\path{doi:10.1103/PhysRevLett.55.1418}}.

\bibitem{hybertsen_electron_1986}
M.~S. Hybertsen, S.~G. Louie, Electron correlation in semiconductors and insulators: {{Band}} gaps and quasiparticle energies, Phys. Rev. B 34~(8) (1986) 5390--5413.
\newblock \href {https://doi.org/10.1103/PhysRevB.34.5390} {\path{doi:10.1103/PhysRevB.34.5390}}.

\bibitem{onida_electronic_2002}
G.~Onida, L.~Reining, A.~Rubio, Electronic excitations: Density-functional versus many-body {{Green}}'s-function approaches, Rev. Mod. Phys. 74~(2) (2002) 601--659.
\newblock \href {https://doi.org/10.1103/RevModPhys.74.601} {\path{doi:10.1103/RevModPhys.74.601}}.

\bibitem{rohlfing_electronhole_1998}
M.~Rohlfing, S.~G. Louie, Electron-{{Hole Excitations}} in {{Semiconductors}} and {{Insulators}}, Phys. Rev. Lett. 81~(11) (1998) 2312--2315.
\newblock \href {https://doi.org/10.1103/PhysRevLett.81.2312} {\path{doi:10.1103/PhysRevLett.81.2312}}.

\bibitem{rohlfing_electronhole_2000}
M.~Rohlfing, S.~G. Louie, Electron-hole excitations and optical spectra from first principles, Phys. Rev. B 62~(8) (2000) 4927--4944.
\newblock \href {https://doi.org/10.1103/PhysRevB.62.4927} {\path{doi:10.1103/PhysRevB.62.4927}}.

\bibitem{kummel_orbitaldependent_2008}
S.~K{\"u}mmel, L.~Kronik, Orbital-dependent density functionals: {{Theory}} and applications, Rev. Mod. Phys. 80~(1) (2008) 3--60.
\newblock \href {https://doi.org/10.1103/RevModPhys.80.3} {\path{doi:10.1103/RevModPhys.80.3}}.

\bibitem{borlido_exchangecorrelation_2020}
P.~Borlido, J.~Schmidt, A.~W. Huran, F.~Tran, M.~A.~L. Marques, S.~Botti, Exchange-correlation functionals for band gaps of solids: Benchmark, reparametrization and machine learning, npj Comput Mater 6~(1) (2020) 96.
\newblock \href {https://doi.org/10.1038/s41524-020-00360-0} {\path{doi:10.1038/s41524-020-00360-0}}.

\bibitem{bryenton_delocalization_2023}
K.~R. Bryenton, A.~A. Adeleke, S.~G. Dale, E.~R. Johnson, Delocalization error: {{The}} greatest outstanding challenge in density-functional theory, WIREs Comput Mol Sci 13~(2) (2023) e1631.
\newblock \href {https://doi.org/10.1002/wcms.1631} {\path{doi:10.1002/wcms.1631}}.

\bibitem{perdew_physical_1983}
J.~P. Perdew, M.~Levy, Physical {{Content}} of the {{Exact Kohn-Sham Orbital Energies}}: {{Band Gaps}} and {{Derivative Discontinuities}}, Phys. Rev. Lett. 51~(20) (1983) 1884--1887.
\newblock \href {https://doi.org/10.1103/PhysRevLett.51.1884} {\path{doi:10.1103/PhysRevLett.51.1884}}.

\bibitem{sham_densityfunctional_1983}
L.~J. Sham, M.~Schl{\"u}ter, Density-{{Functional Theory}} of the {{Energy Gap}}, Phys. Rev. Lett. 51~(20) (1983) 1888--1891.
\newblock \href {https://doi.org/10.1103/PhysRevLett.51.1888} {\path{doi:10.1103/PhysRevLett.51.1888}}.

\bibitem{ullrich_chapter_2016}
C.~A. Ullrich, Chapter 2 of {{Time-dependent}} Density-Functional Theory: Concepts and Applications, reprinted, with corrections Edition, Oxford Graduate Texts, Oxford University Press, Oxford New York, 2016.

\bibitem{perdew_densityfunctional_1982}
J.~P. Perdew, R.~G. Parr, M.~Levy, J.~L. Balduz, Density-{{Functional Theory}} for {{Fractional Particle Number}}: {{Derivative Discontinuities}} of the {{Energy}}, Phys. Rev. Lett. 49~(23) (1982) 1691--1694.
\newblock \href {https://doi.org/10.1103/PhysRevLett.49.1691} {\path{doi:10.1103/PhysRevLett.49.1691}}.

\bibitem{bylander_good_1990}
D.~M. Bylander, L.~Kleinman, Good semiconductor band gaps with a modified local-density approximation, Phys. Rev. B 41~(11) (1990) 7868--7871.
\newblock \href {https://doi.org/10.1103/PhysRevB.41.7868} {\path{doi:10.1103/PhysRevB.41.7868}}.

\bibitem{geller_computational_2001}
C.~B. Geller, W.~Wolf, S.~Picozzi, A.~Continenza, R.~Asahi, W.~Mannstadt, A.~J. Freeman, E.~Wimmer, Computational band-structure engineering of {{III}}--{{V}} semiconductor alloys, Appl. Phys. Lett. 79~(3) (2001) 368--370.
\newblock \href {https://doi.org/10.1063/1.1383282} {\path{doi:10.1063/1.1383282}}.

\bibitem{heyd_energy_2005}
J.~Heyd, J.~E. Peralta, G.~E. Scuseria, R.~L. Martin, Energy band gaps and lattice parameters evaluated with the {{Heyd-Scuseria-Ernzerhof}} screened hybrid functional, J. Chem. Phys. 123~(17) (2005) 174101.
\newblock \href {https://doi.org/10.1063/1.2085170} {\path{doi:10.1063/1.2085170}}.

\bibitem{cococcioni_linear_2005}
M.~Cococcioni, S.~De~Gironcoli, Linear response approach to the calculation of the effective interaction parameters in the {{LDA}} + {{U}} method, Phys. Rev. B 71~(3) (2005) 035105.
\newblock \href {https://doi.org/10.1103/PhysRevB.71.035105} {\path{doi:10.1103/PhysRevB.71.035105}}.

\bibitem{anisimov_transition_2005}
V.~I. Anisimov, A.~V. Kozhevnikov, Transition state method and {{Wannier}} functions, Phys. Rev. B 72~(7) (2005) 075125.
\newblock \href {https://doi.org/10.1103/PhysRevB.72.075125} {\path{doi:10.1103/PhysRevB.72.075125}}.

\bibitem{ferreira_approximation_2008}
L.~G. Ferreira, M.~Marques, L.~K. Teles, Approximation to density functional theory for the calculation of band gaps of semiconductors, Phys. Rev. B 78~(12) (2008) 125116.
\newblock \href {https://doi.org/10.1103/PhysRevB.78.125116} {\path{doi:10.1103/PhysRevB.78.125116}}.

\bibitem{shimazaki_band_2008}
T.~Shimazaki, Y.~Asai, Band structure calculations based on screened {{Fock}} exchange method, Chemical Physics Letters 466~(1-3) (2008) 91--94.
\newblock \href {https://doi.org/10.1016/j.cplett.2008.10.012} {\path{doi:10.1016/j.cplett.2008.10.012}}.

\bibitem{zhao_calculation_2009}
Y.~Zhao, D.~G. Truhlar, Calculation of semiconductor band gaps with the {{M06-L}} density functional, J. Chem. Phys. 130~(7) (2009) 074103.
\newblock \href {https://doi.org/10.1063/1.3076922} {\path{doi:10.1063/1.3076922}}.

\bibitem{tran_accurate_2009}
F.~Tran, P.~Blaha, Accurate {{Band Gaps}} of {{Semiconductors}} and {{Insulators}} with a {{Semilocal Exchange-Correlation Potential}}, Phys. Rev. Lett. 102~(22) (2009) 226401.
\newblock \href {https://doi.org/10.1103/PhysRevLett.102.226401} {\path{doi:10.1103/PhysRevLett.102.226401}}.

\bibitem{chan_efficient_2010}
M.~K.~Y. Chan, G.~Ceder, Efficient {{Band Gap Prediction}} for {{Solids}}, Phys. Rev. Lett. 105~(19) (2010) 196403.
\newblock \href {https://doi.org/10.1103/PhysRevLett.105.196403} {\path{doi:10.1103/PhysRevLett.105.196403}}.

\bibitem{dabo_firstprinciples_2010}
I.~Dabo, E.~Canc{\`e}s, Y.~L. Li, N.~Marzari, Towards {{First-principles Electrochemistry}} (Sep. 2010).
\newblock \href {http://arxiv.org/abs/0901.0096} {\path{arXiv:0901.0096}}, \href {https://doi.org/10.48550/arXiv.0901.0096} {\path{doi:10.48550/arXiv.0901.0096}}.

\bibitem{dabo_koopmans_2010}
I.~Dabo, A.~Ferretti, N.~Poilvert, Y.~Li, N.~Marzari, M.~Cococcioni, Koopmans' condition for density-functional theory, Phys. Rev. B 82~(11) (2010) 115121.
\newblock \href {https://doi.org/10.1103/PhysRevB.82.115121} {\path{doi:10.1103/PhysRevB.82.115121}}.

\bibitem{marques_densitybased_2011}
M.~A.~L. Marques, J.~Vidal, M.~J.~T. Oliveira, L.~Reining, S.~Botti, Density-based mixing parameter for hybrid functionals, Phys. Rev. B 83~(3) (2011) 035119.
\newblock \href {https://doi.org/10.1103/PhysRevB.83.035119} {\path{doi:10.1103/PhysRevB.83.035119}}.

\bibitem{refaely-abramson_gap_2013}
S.~{Refaely-Abramson}, S.~Sharifzadeh, M.~Jain, R.~Baer, J.~B. Neaton, L.~Kronik, Gap renormalization of molecular crystals from density-functional theory, Phys. Rev. B 88~(8) (2013) 081204.
\newblock \href {https://doi.org/10.1103/PhysRevB.88.081204} {\path{doi:10.1103/PhysRevB.88.081204}}.

\bibitem{skone_selfconsistent_2014}
J.~H. Skone, M.~Govoni, G.~Galli, Self-consistent hybrid functional for condensed systems, Phys. Rev. B 89~(19) (2014) 195112.
\newblock \href {https://doi.org/10.1103/PhysRevB.89.195112} {\path{doi:10.1103/PhysRevB.89.195112}}.

\bibitem{borghi_koopmanscompliant_2014}
G.~Borghi, A.~Ferretti, N.~L. Nguyen, I.~Dabo, N.~Marzari, Koopmans-compliant functionals and their performance against reference molecular data, Phys. Rev. B 90~(7) (2014) 075135.
\newblock \href {https://doi.org/10.1103/PhysRevB.90.075135} {\path{doi:10.1103/PhysRevB.90.075135}}.

\bibitem{gorling_exchangecorrelation_2015}
A.~G{\"o}rling, Exchange-correlation potentials with proper discontinuities for physically meaningful {{Kohn-Sham}} eigenvalues and band structures, Phys. Rev. B 91~(24) (2015) 245120.
\newblock \href {https://doi.org/10.1103/PhysRevB.91.245120} {\path{doi:10.1103/PhysRevB.91.245120}}.

\bibitem{skone_nonempirical_2016}
J.~H. Skone, M.~Govoni, G.~Galli, Nonempirical range-separated hybrid functionals for solids and molecules, Phys. Rev. B 93~(23) (2016) 235106.
\newblock \href {https://doi.org/10.1103/PhysRevB.93.235106} {\path{doi:10.1103/PhysRevB.93.235106}}.

\bibitem{ma_using_2016}
J.~Ma, L.-W. Wang, Using {{Wannier}} functions to improve solid band gap predictions in density functional theory, Sci Rep 6~(1) (2016) 24924.
\newblock \href {https://doi.org/10.1038/srep24924} {\path{doi:10.1038/srep24924}}.

\bibitem{weng_wannier_2017}
M.~Weng, S.~Li, J.~Ma, J.~Zheng, F.~Pan, L.-W. Wang, Wannier {{Koopman}} method calculations of the band gaps of alkali halides, Appl. Phys. Lett. 111~(5) (2017) 054101.
\newblock \href {https://doi.org/10.1063/1.4996743} {\path{doi:10.1063/1.4996743}}.

\bibitem{perdew_understanding_2017}
J.~P. Perdew, W.~Yang, K.~Burke, Z.~Yang, E.~K.~U. Gross, M.~Scheffler, G.~E. Scuseria, T.~M. Henderson, I.~Y. Zhang, A.~Ruzsinszky, H.~Peng, J.~Sun, E.~Trushin, A.~G{\"o}rling, Understanding band gaps of solids in generalized {{Kohn}}--{{Sham}} theory, Proc. Natl. Acad. Sci. U.S.A. 114~(11) (2017) 2801--2806.
\newblock \href {https://doi.org/10.1073/pnas.1621352114} {\path{doi:10.1073/pnas.1621352114}}.

\bibitem{verma_hle17_2017}
P.~Verma, D.~G. Truhlar, {{HLE17}}: {{An Improved Local Exchange}}--{{Correlation Functional}} for {{Computing Semiconductor Band Gaps}} and {{Molecular Excitation Energies}}, J. Phys. Chem. C 121~(13) (2017) 7144--7154.
\newblock \href {https://doi.org/10.1021/acs.jpcc.7b01066} {\path{doi:10.1021/acs.jpcc.7b01066}}.

\bibitem{nguyen_koopmanscompliant_2018}
N.~L. Nguyen, N.~Colonna, A.~Ferretti, N.~Marzari, Koopmans-{{Compliant Spectral Functionals}} for {{Extended Systems}}, Phys. Rev. X 8~(2) (2018) 021051.
\newblock \href {https://doi.org/10.1103/PhysRevX.8.021051} {\path{doi:10.1103/PhysRevX.8.021051}}.

\bibitem{cui_doubly_2018}
Z.-H. Cui, Y.-C. Wang, M.-Y. Zhang, X.~Xu, H.~Jiang, Doubly {{Screened Hybrid Functional}}: {{An Accurate First-Principles Approach}} for {{Both Narrow-}} and {{Wide-Gap Semiconductors}}, J. Phys. Chem. Lett. 9~(9) (2018) 2338--2345.
\newblock \href {https://doi.org/10.1021/acs.jpclett.8b00919} {\path{doi:10.1021/acs.jpclett.8b00919}}.

\bibitem{chen_nonempirical_2018}
W.~Chen, G.~Miceli, G.-M. Rignanese, A.~Pasquarello, Nonempirical dielectric-dependent hybrid functional with range separation for semiconductors and insulators, Phys. Rev. Materials 2~(7) (2018) 073803.
\newblock \href {https://doi.org/10.1103/PhysRevMaterials.2.073803} {\path{doi:10.1103/PhysRevMaterials.2.073803}}.

\bibitem{miceli_nonempirical_2018}
G.~Miceli, W.~Chen, I.~Reshetnyak, A.~Pasquarello, Nonempirical hybrid functionals for band gaps and polaronic distortions in solids, Phys. Rev. B 97~(12) (2018) 121112.
\newblock \href {https://doi.org/10.1103/PhysRevB.97.121112} {\path{doi:10.1103/PhysRevB.97.121112}}.

\bibitem{colonna_screening_2018}
N.~Colonna, N.~L. Nguyen, A.~Ferretti, N.~Marzari, Screening in {{Orbital-Density-Dependent Functionals}}, J. Chem. Theory Comput. 14~(5) (2018) 2549--2557.
\newblock \href {https://doi.org/10.1021/acs.jctc.7b01116} {\path{doi:10.1021/acs.jctc.7b01116}}.

\bibitem{bischoff_nonempirical_2019}
T.~Bischoff, J.~Wiktor, W.~Chen, A.~Pasquarello, Nonempirical hybrid functionals for band gaps of inorganic metal-halide perovskites, Phys. Rev. Materials 3~(12) (2019) 123802.
\newblock \href {https://doi.org/10.1103/PhysRevMaterials.3.123802} {\path{doi:10.1103/PhysRevMaterials.3.123802}}.

\bibitem{aschebrock_ultranonlocality_2019}
T.~Aschebrock, S.~K{\"u}mmel, Ultranonlocality and accurate band gaps from a meta-generalized gradient approximation, Phys. Rev. Research 1~(3) (2019) 033082.
\newblock \href {https://doi.org/10.1103/PhysRevResearch.1.033082} {\path{doi:10.1103/PhysRevResearch.1.033082}}.

\bibitem{colonna_koopmanscompliant_2019}
N.~Colonna, N.~L. Nguyen, A.~Ferretti, N.~Marzari, Koopmans-{{Compliant Functionals}} and {{Potentials}} and {{Their Application}} to the {{GW100 Test Set}}, J. Chem. Theory Comput. 15~(3) (2019) 1905--1914.
\newblock \href {https://doi.org/10.1021/acs.jctc.8b00976} {\path{doi:10.1021/acs.jctc.8b00976}}.

\bibitem{weng_wannier_2020}
M.~Weng, F.~Pan, L.-W. Wang, Wannier--{{Koopmans}} method calculations for transition metal oxide band gaps, npj Comput Mater 6~(1) (2020) 33.
\newblock \href {https://doi.org/10.1038/s41524-020-0302-0} {\path{doi:10.1038/s41524-020-0302-0}}.

\bibitem{cipriano_band_2020}
L.~A. Cipriano, G.~Di~Liberto, S.~Tosoni, G.~Pacchioni, Band {{Gap}} in {{Magnetic Insulators}} from a {{Charge Transition Level Approach}}, J. Chem. Theory Comput. 16~(6) (2020) 3786--3798.
\newblock \href {https://doi.org/10.1021/acs.jctc.0c00134} {\path{doi:10.1021/acs.jctc.0c00134}}.

\bibitem{tancogne-dejean_parameterfree_2020}
N.~{Tancogne-Dejean}, A.~Rubio, Parameter-free hybridlike functional based on an extended {{Hubbard}} model: {{DFT}} + {{U}} + {{V}}, Phys. Rev. B 102~(15) (2020) 155117.
\newblock \href {https://doi.org/10.1103/PhysRevB.102.155117} {\path{doi:10.1103/PhysRevB.102.155117}}.

\bibitem{lee_firstprinciples_2020}
S.-H. Lee, Y.-W. Son, First-principles approach with a pseudohybrid density functional for extended {{Hubbard}} interactions, Phys. Rev. Research 2~(4) (2020) 043410.
\newblock \href {https://doi.org/10.1103/PhysRevResearch.2.043410} {\path{doi:10.1103/PhysRevResearch.2.043410}}.

\bibitem{lorke_koopmanscompliant_2020}
M.~Lorke, P.~De{\'a}k, T.~Frauenheim, Koopmans-compliant screened exchange potential with correct asymptotic behavior for semiconductors, Phys. Rev. B 102~(23) (2020) 235168.
\newblock \href {https://doi.org/10.1103/PhysRevB.102.235168} {\path{doi:10.1103/PhysRevB.102.235168}}.

\bibitem{wing_band_2021}
D.~Wing, G.~Ohad, J.~B. Haber, M.~R. Filip, S.~E. Gant, J.~B. Neaton, L.~Kronik, Band gaps of crystalline solids from {{Wannier-localization}}--based optimal tuning of a screened range-separated hybrid functional, Proc. Natl. Acad. Sci. U.S.A. 118~(34) (2021) e2104556118.
\newblock \href {https://doi.org/10.1073/pnas.2104556118} {\path{doi:10.1073/pnas.2104556118}}.

\bibitem{colonna_koopmans_2022}
N.~Colonna, R.~De~Gennaro, E.~Linscott, N.~Marzari, Koopmans {{Spectral Functionals}} in {{Periodic Boundary Conditions}}, J. Chem. Theory Comput. 18~(9) (2022) 5435--5448.
\newblock \href {https://doi.org/10.1021/acs.jctc.2c00161} {\path{doi:10.1021/acs.jctc.2c00161}}.

\bibitem{yang_oneshot_2022}
J.~Yang, S.~Falletta, A.~Pasquarello, One-{{Shot Approach}} for {{Enforcing Piecewise Linearity}} on {{Hybrid Functionals}}: {{Application}} to {{Band Gap Predictions}}, J. Phys. Chem. Lett. 13~(13) (2022) 3066--3071.
\newblock \href {https://doi.org/10.1021/acs.jpclett.2c00414} {\path{doi:10.1021/acs.jpclett.2c00414}}.

\bibitem{yang_rangeseparated_2023}
J.~Yang, S.~Falletta, A.~Pasquarello, Range-separated hybrid functionals for accurate prediction of band gaps of extended systems, npj Comput Mater 9~(1) (2023) 1--9.
\newblock \href {https://doi.org/10.1038/s41524-023-01064-x} {\path{doi:10.1038/s41524-023-01064-x}}.

\bibitem{zhan_nonempirical_2023}
J.~Zhan, M.~Govoni, G.~Galli, Nonempirical {{Range-Separated Hybrid Functional}} with {{Spatially Dependent Screened Exchange}}, J. Chem. Theory Comput. 19~(17) (2023) 5851--5862.
\newblock \href {https://doi.org/10.1021/acs.jctc.3c00580} {\path{doi:10.1021/acs.jctc.3c00580}}.

\bibitem{degennaro_blochs_2022}
R.~De~Gennaro, N.~Colonna, E.~Linscott, N.~Marzari, Bloch's theorem in orbital-density-dependent functionals: {{Band}} structures from {{Koopmans}} spectral functionals, Phys. Rev. B 106~(3) (2022) 035106.
\newblock \href {https://doi.org/10.1103/PhysRevB.106.035106} {\path{doi:10.1103/PhysRevB.106.035106}}.

\bibitem{ohad_optical_2023}
G.~Ohad, S.~E. Gant, D.~Wing, J.~B. Haber, M.~{Camarasa-G{\'o}mez}, F.~Sagredo, M.~R. Filip, J.~B. Neaton, L.~Kronik, Optical absorption spectra of metal oxides from time-dependent density functional theory and many-body perturbation theory based on optimally-tuned hybrid functionals, Phys. Rev. Materials 7~(12) (2023) 123803.
\newblock \href {https://doi.org/10.1103/PhysRevMaterials.7.123803} {\path{doi:10.1103/PhysRevMaterials.7.123803}}.

\bibitem{linscott_koopmans_2023}
E.~B. Linscott, N.~Colonna, R.~De~Gennaro, N.~L. Nguyen, G.~Borghi, A.~Ferretti, I.~Dabo, N.~Marzari, Koopmans : {{An Open-Source Package}} for {{Accurately}} and {{Efficiently Predicting Spectral Properties}} with {{Koopmans Functionals}}, J. Chem. Theory Comput. 19~(20) (2023) 7097--7111.
\newblock \href {https://doi.org/10.1021/acs.jctc.3c00652} {\path{doi:10.1021/acs.jctc.3c00652}}.

\bibitem{camarasa-gomez_excitations_2024}
M.~{Camarasa-G{\'o}mez}, S.~E. Gant, G.~Ohad, J.~B. Neaton, A.~Ramasubramaniam, L.~Kronik, Excitations in layered materials from a non-empirical {{Wannier-localized}} optimally-tuned screened range-separated hybrid functional, npj Comput Mater 10~(1) (2024) 288.
\newblock \href {https://doi.org/10.1038/s41524-024-01478-1} {\path{doi:10.1038/s41524-024-01478-1}}.

\bibitem{schubert_predicting_2024}
Y.~Schubert, S.~Luber, N.~Marzari, E.~Linscott, Predicting electronic screening for fast {{Koopmans}} spectral functional calculations, npj Comput Mater 10~(1) (2024) 299.
\newblock \href {https://doi.org/10.1038/s41524-024-01484-3} {\path{doi:10.1038/s41524-024-01484-3}}.

\bibitem{zhan_dielectricdependent_2025}
J.~Zhan, M.~Govoni, G.~Galli, Dielectric-{{Dependent Range-Separated Hybrid Functional Calculations}} for {{Metal Oxides}} (Feb. 2025).
\newblock \href {http://arxiv.org/abs/2502.13035} {\path{arXiv:2502.13035}}, \href {https://doi.org/10.48550/arXiv.2502.13035} {\path{doi:10.48550/arXiv.2502.13035}}.

\bibitem{seidl_generalized_1996}
A.~Seidl, A.~G{\"o}rling, P.~Vogl, J.~A. Majewski, M.~Levy, Generalized {{Kohn-Sham}} schemes and the band-gap problem, Phys. Rev. B 53~(7) (1996) 3764--3774.
\newblock \href {https://doi.org/10.1103/PhysRevB.53.3764} {\path{doi:10.1103/PhysRevB.53.3764}}.

\bibitem{baer_timedependent_2018}
R.~Baer, L.~Kronik, Time-dependent generalized {{Kohn}}--{{Sham}} theory, Eur. Phys. J. B 91~(7) (2018) 170.
\newblock \href {https://doi.org/10.1140/epjb/e2018-90103-0} {\path{doi:10.1140/epjb/e2018-90103-0}}.

\bibitem{garrick_exact_2020}
R.~Garrick, A.~Natan, T.~Gould, L.~Kronik, Exact {{Generalized Kohn-Sham Theory}} for {{Hybrid Functionals}}, Phys. Rev. X 10~(2) (2020) 021040.
\newblock \href {https://doi.org/10.1103/PhysRevX.10.021040} {\path{doi:10.1103/PhysRevX.10.021040}}.

\bibitem{refaely-abramson_solidstate_2015}
S.~{Refaely-Abramson}, M.~Jain, S.~Sharifzadeh, J.~B. Neaton, L.~Kronik, Solid-state optical absorption from optimally tuned time-dependent range-separated hybrid density functional theory, Phys. Rev. B 92~(8) (2015) 081204.
\newblock \href {https://doi.org/10.1103/PhysRevB.92.081204} {\path{doi:10.1103/PhysRevB.92.081204}}.

\bibitem{sagredo_electronic_2024}
F.~Sagredo, S.~E. Gant, G.~Ohad, J.~B. Haber, M.~R. Filip, L.~Kronik, J.~B. Neaton, Electronic structure and optical properties of halide double perovskites from a {{Wannier-localized}} optimally-tuned screened range-separated hybrid functional, Phys. Rev. Materials 8~(10) (2024) 105401.
\newblock \href {https://doi.org/10.1103/PhysRevMaterials.8.105401} {\path{doi:10.1103/PhysRevMaterials.8.105401}}.

\bibitem{ohad_band_2022}
G.~Ohad, D.~Wing, S.~E. Gant, A.~V. Cohen, J.~B. Haber, F.~Sagredo, M.~R. Filip, J.~B. Neaton, L.~Kronik, Band gaps of halide perovskites from a {{Wannier-localized}} optimally tuned screened range-separated hybrid functional, Phys. Rev. Materials 6~(10) (2022) 104606.
\newblock \href {https://doi.org/10.1103/PhysRevMaterials.6.104606} {\path{doi:10.1103/PhysRevMaterials.6.104606}}.

\bibitem{florio_resolving_2025}
F.~Florio, M.~{Camarasa-G{\'o}mez}, G.~Ohad, D.~Naveh, L.~Kronik, A.~Ramasubramaniam, Resolving contradictory estimates of band gaps of bulk {{PdSe}}{\textsubscript{2}}: {{A Wannier-localized}} optimally-tuned screened range-separated hybrid density functional theory study, Appl. Phys. Lett. 126~(14) (Apr. 2025).
\newblock \href {https://doi.org/10.1063/5.0260649} {\path{doi:10.1063/5.0260649}}.

\bibitem{ohad_nonempirical_2024}
G.~Ohad, M.~Hartstein, T.~Gould, J.~B. Neaton, L.~Kronik, Nonempirical {{Prediction}} of the {{Length-Dependent Ionization Potential}} in {{Molecular Chains}}, J. Chem. Theory Comput. (2024) acs.jctc.4c00847\href {https://doi.org/10.1021/acs.jctc.4c00847} {\path{doi:10.1021/acs.jctc.4c00847}}.

\bibitem{sagredo_reliability_2025}
F.~Sagredo, M.~{Camarasa-G{\'o}mez}, F.~Ricci, A.~Champagne, L.~Kronik, J.~B. Neaton, The {{Reliability}} of {{Hybrid Functionals}} for {{Accurate Fundamental}} and {{Optical Gap Prediction}} of {{Bulk Solids}} and {{Surfaces}}, J. Chem. Theory Comput. 21~(10) (2025) 5009--5015.
\newblock \href {https://doi.org/10.1021/acs.jctc.5c00160} {\path{doi:10.1021/acs.jctc.5c00160}}.

\bibitem{ke_accurate_2025}
S.~Ke, S.~E. Gant, L.~Kronik, J.~B. Neaton, Accurate point defect energy levels from non-empirical screened range-separated hybrid functionals: {{The}} case of native vacancies in {{ZnO}}, Phys. Rev. Materials 9~(5) (2025) 053806.
\newblock \href {https://doi.org/10.1103/PhysRevMaterials.9.053806} {\path{doi:10.1103/PhysRevMaterials.9.053806}}.

\bibitem{gant_optimally_2022}
S.~E. Gant, J.~B. Haber, M.~R. Filip, F.~Sagredo, D.~Wing, G.~Ohad, L.~Kronik, J.~B. Neaton, Optimally tuned starting point for single-shot {{{\emph{GW}}}} calculations of solids, Phys. Rev. Materials 6~(5) (2022) 053802.
\newblock \href {https://doi.org/10.1103/PhysRevMaterials.6.053802} {\path{doi:10.1103/PhysRevMaterials.6.053802}}.

\bibitem{marzari_maximally_1997}
N.~Marzari, D.~Vanderbilt, Maximally localized generalized {{Wannier}} functions for composite energy bands, Phys. Rev. B 56~(20) (1997) 12847--12865.
\newblock \href {https://doi.org/10.1103/PhysRevB.56.12847} {\path{doi:10.1103/PhysRevB.56.12847}}.

\bibitem{souza_maximally_2001}
I.~Souza, N.~Marzari, D.~Vanderbilt, Maximally localized {{Wannier}} functions for entangled energy bands, Phys. Rev. B 65~(3) (2001) 035109.
\newblock \href {https://doi.org/10.1103/PhysRevB.65.035109} {\path{doi:10.1103/PhysRevB.65.035109}}.

\bibitem{kronik_excitedstate_2016}
L.~Kronik, J.~B. Neaton, Excited-{{State Properties}} of {{Molecular Solids}} from {{First Principles}}, Annu. Rev. Phys. Chem. 67~(1) (2016) 587--616.
\newblock \href {https://doi.org/10.1146/annurev-physchem-040214-121351} {\path{doi:10.1146/annurev-physchem-040214-121351}}.

\bibitem{becke_new_1993}
A.~D. Becke, A new mixing of {{Hartree}}--{{Fock}} and local density-functional theories, The Journal of Chemical Physics 98~(2) (1993) 1372--1377.
\newblock \href {https://doi.org/10.1063/1.464304} {\path{doi:10.1063/1.464304}}.

\bibitem{perdew_rationale_1996}
J.~P. Perdew, M.~Ernzerhof, K.~Burke, Rationale for mixing exact exchange with density functional approximations, The Journal of Chemical Physics 105~(22) (1996) 9982--9985.
\newblock \href {https://doi.org/10.1063/1.472933} {\path{doi:10.1063/1.472933}}.

\bibitem{yanai_new_2004}
T.~Yanai, D.~P. Tew, N.~C. Handy, A new hybrid exchange--correlation functional using the {{Coulomb-attenuating}} method ({{CAM-B3LYP}}), Chemical Physics Letters 393~(1) (2004) 51--57.
\newblock \href {https://doi.org/10.1016/j.cplett.2004.06.011} {\path{doi:10.1016/j.cplett.2004.06.011}}.

\bibitem{refaely-abramson_quasiparticle_2012}
S.~{Refaely-Abramson}, S.~Sharifzadeh, N.~Govind, J.~Autschbach, J.~B. Neaton, R.~Baer, L.~Kronik, Quasiparticle {{Spectra}} from a {{Nonempirical Optimally Tuned Range-Separated Hybrid Density Functional}}, Phys. Rev. Lett. 109~(22) (2012) 226405.
\newblock \href {https://doi.org/10.1103/PhysRevLett.109.226405} {\path{doi:10.1103/PhysRevLett.109.226405}}.

\bibitem{srebro_does_2012}
M.~Srebro, J.~Autschbach, Does a {{Molecule-Specific Density Functional Give}} an {{Accurate Electron Density}}? {{The Challenging Case}} of the {{CuCl Electric Field Gradient}}, J. Phys. Chem. Lett. 3~(5) (2012) 576--581.
\newblock \href {https://doi.org/10.1021/jz201685r} {\path{doi:10.1021/jz201685r}}.

\bibitem{ohad_foundations_2025}
G.~Ohad, M.~{Camarasa-G{\'o}mez}, J.~B. Neaton, A.~Ramasubramaniam, T.~Gould, L.~Kronik, Foundations of the ionization potential condition for localized electron removal in density functional theory (May 2025).
\newblock \href {http://arxiv.org/abs/2506.00629} {\path{arXiv:2506.00629}}, \href {https://doi.org/10.48550/arXiv.2506.00629} {\path{doi:10.48550/arXiv.2506.00629}}.

\bibitem{levy_exact_1984}
M.~Levy, J.~P. Perdew, V.~Sahni, Exact differential equation for the density and ionization energy of a many-particle system, Phys. Rev. A 30~(5) (1984) 2745--2748.
\newblock \href {https://doi.org/10.1103/PhysRevA.30.2745} {\path{doi:10.1103/PhysRevA.30.2745}}.

\bibitem{almbladh_exact_1985}
C.-O. Almbladh, U.~Von~Barth, Exact results for the charge and spin densities, exchange-correlation potentials, and density-functional eigenvalues, Phys. Rev. B 31~(6) (1985) 3231--3244.
\newblock \href {https://doi.org/10.1103/PhysRevB.31.3231} {\path{doi:10.1103/PhysRevB.31.3231}}.

\bibitem{leslie_energy_1985}
M.~Leslie, N.~J. Gillan, The energy and elastic dipole tensor of defects in ionic crystals calculated by the supercell method, J. Phys. C: Solid State Phys. 18~(5) (1985) 973.
\newblock \href {https://doi.org/10.1088/0022-3719/18/5/005} {\path{doi:10.1088/0022-3719/18/5/005}}.

\bibitem{makov_periodic_1995}
G.~Makov, M.~C. Payne, Periodic boundary conditions in {\emph{ab initio}} calculations, Phys. Rev. B 51~(7) (1995) 4014--4022.
\newblock \href {https://doi.org/10.1103/PhysRevB.51.4014} {\path{doi:10.1103/PhysRevB.51.4014}}.

\bibitem{komsa_finitesize_2012}
H.-P. Komsa, T.~T. Rantala, A.~Pasquarello, Finite-size supercell correction schemes for charged defect calculations, Phys. Rev. B 86~(4) (2012) 045112.
\newblock \href {https://doi.org/10.1103/PhysRevB.86.045112} {\path{doi:10.1103/PhysRevB.86.045112}}.

\bibitem{rurali_theory_2009}
R.~Rurali, X.~Cartoix{\`a}, Theory of {{Defects}} in {{One-Dimensional Systems}}: {{Application}} to {{Al-Catalyzed Si Nanowires}}, Nano Lett. 9~(3) (2009) 975--979.
\newblock \href {https://doi.org/10.1021/nl802847p} {\path{doi:10.1021/nl802847p}}.

\bibitem{lany_assessment_2008}
S.~Lany, A.~Zunger, Assessment of correction methods for the band-gap problem and for finite-size effects in supercell defect calculations: {{Case}} studies for {{ZnO}} and {{GaAs}}, Phys. Rev. B 78~(23) (2008) 235104.
\newblock \href {https://doi.org/10.1103/PhysRevB.78.235104} {\path{doi:10.1103/PhysRevB.78.235104}}.

\bibitem{ramasubramaniam_transferable_2019}
A.~Ramasubramaniam, D.~Wing, L.~Kronik, Transferable screened range-separated hybrids for layered materials: {{The}} cases of {{MoS}}{\textsubscript{2}} and h-{{BN}}, Phys. Rev. Materials 3~(8) (2019) 084007.
\newblock \href {https://doi.org/10.1103/PhysRevMaterials.3.084007} {\path{doi:10.1103/PhysRevMaterials.3.084007}}.

\bibitem{camarasa-gomez_transferable_2023}
M.~{Camarasa-G{\'o}mez}, A.~Ramasubramaniam, J.~B. Neaton, L.~Kronik, Transferable screened range-separated hybrid functionals for electronic and optical properties of van der {{Waals}} materials, Phys. Rev. Mater. 7~(10) (2023) 104001.
\newblock \href {https://doi.org/10.1103/PhysRevMaterials.7.104001} {\path{doi:10.1103/PhysRevMaterials.7.104001}}.

\bibitem{ong_python_2013}
S.~P. Ong, W.~D. Richards, A.~Jain, G.~Hautier, M.~Kocher, S.~Cholia, D.~Gunter, V.~L. Chevrier, K.~A. Persson, G.~Ceder, Python {{Materials Genomics}} (pymatgen): {{A}} robust, open-source python library for materials analysis, Computational Materials Science 68 (2013) 314--319.
\newblock \href {https://doi.org/10.1016/j.commatsci.2012.10.028} {\path{doi:10.1016/j.commatsci.2012.10.028}}.

\bibitem{perdew_generalized_1996}
J.~P. Perdew, K.~Burke, M.~Ernzerhof, Generalized {{Gradient Approximation Made Simple}}, Phys. Rev. Lett. 77~(18) (1996) 3865--3868.
\newblock \href {https://doi.org/10.1103/PhysRevLett.77.3865} {\path{doi:10.1103/PhysRevLett.77.3865}}.

\bibitem{krukau_influence_2006}
A.~V. Krukau, O.~A. Vydrov, A.~F. Izmaylov, G.~E. Scuseria, Influence of the exchange screening parameter on the performance of screened hybrid functionals, The Journal of Chemical Physics 125~(22) (2006) 224106.
\newblock \href {https://doi.org/10.1063/1.2404663} {\path{doi:10.1063/1.2404663}}.

\bibitem{mostofi_updated_2014}
A.~A. Mostofi, J.~R. Yates, G.~Pizzi, Y.-S. Lee, I.~Souza, D.~Vanderbilt, N.~Marzari, An updated version of wannier90: {{A}} tool for obtaining maximally-localised {{Wannier}} functions, Comput. Phys. Commun. 185~(8) (2014) 2309--2310.
\newblock \href {https://doi.org/10.1016/j.cpc.2014.05.003} {\path{doi:10.1016/j.cpc.2014.05.003}}.

\bibitem{vitale_automated_2020}
V.~Vitale, G.~Pizzi, A.~Marrazzo, J.~R. Yates, N.~Marzari, A.~A. Mostofi, Automated high-throughput {{Wannierisation}}, npj Comput Mater 6~(1) (2020) 1--18.
\newblock \href {https://doi.org/10.1038/s41524-020-0312-y} {\path{doi:10.1038/s41524-020-0312-y}}.

\bibitem{rosen_jobflow_2024}
A.~S. Rosen, M.~Gallant, J.~George, J.~Riebesell, H.~Sahasrabuddhe, J.-X. Shen, M.~Wen, M.~L. Evans, G.~Petretto, D.~Waroquiers, G.-M. Rignanese, K.~A. Persson, A.~Jain, A.~M. Ganose, Jobflow: {{Computational Workflows Made Simple}}, J. Open Source Softw. 9~(93) (2024) 5995.
\newblock \href {https://doi.org/10.21105/joss.05995} {\path{doi:10.21105/joss.05995}}.

\bibitem{kresse_initio_1993}
G.~Kresse, J.~Hafner, Ab initio molecular dynamics for liquid metals, Phys. Rev. B 47~(1) (1993) 558--561.
\newblock \href {https://doi.org/10.1103/PhysRevB.47.558} {\path{doi:10.1103/PhysRevB.47.558}}.

\bibitem{kresse_initio_1994}
G.~Kresse, J.~Hafner, Ab initio molecular-dynamics simulation of the liquid-metal--amorphous-semiconductor transition in germanium, Phys. Rev. B 49~(20) (1994) 14251--14269.
\newblock \href {https://doi.org/10.1103/PhysRevB.49.14251} {\path{doi:10.1103/PhysRevB.49.14251}}.

\bibitem{kresse_normconserving_1994}
G.~Kresse, J.~Hafner, Norm-conserving and ultrasoft pseudopotentials for first-row and transition elements, J. Phys.: Condens. Matter 6~(40) (1994) 8245.
\newblock \href {https://doi.org/10.1088/0953-8984/6/40/015} {\path{doi:10.1088/0953-8984/6/40/015}}.

\bibitem{kresse_efficiency_1996}
G.~Kresse, J.~Furthm{\"u}ller, Efficiency of ab-initio total energy calculations for metals and semiconductors using a plane-wave basis set, Computational Materials Science 6~(1) (1996) 15--50.
\newblock \href {https://doi.org/10.1016/0927-0256(96)00008-0} {\path{doi:10.1016/0927-0256(96)00008-0}}.

\bibitem{kresse_efficient_1996}
G.~Kresse, J.~Furthm{\"u}ller, Efficient iterative schemes for ab initio total-energy calculations using a plane-wave basis set, Phys. Rev. B 54~(16) (1996) 11169--11186.
\newblock \href {https://doi.org/10.1103/PhysRevB.54.11169} {\path{doi:10.1103/PhysRevB.54.11169}}.

\bibitem{kresse_ultrasoft_1999}
G.~Kresse, D.~Joubert, From ultrasoft pseudopotentials to the projector augmented-wave method, Phys. Rev. B 59~(3) (1999) 1758--1775.
\newblock \href {https://doi.org/10.1103/PhysRevB.59.1758} {\path{doi:10.1103/PhysRevB.59.1758}}.

\bibitem{usuda_allelectron_2002}
M.~Usuda, N.~Hamada, T.~Kotani, M.~{van Schilfgaarde}, All-electron {{{\emph{GW}}}} calculation based on the {{LAPW}} method: {{Application}} to wurtzite {{ZnO}}, Phys. Rev. B 66~(12) (2002) 125101.
\newblock \href {https://doi.org/10.1103/PhysRevB.66.125101} {\path{doi:10.1103/PhysRevB.66.125101}}.

\bibitem{rinke_combining_2005}
P.~Rinke, A.~Qteish, J.~Neugebauer, C.~Freysoldt, M.~Scheffler, Combining {{GW}} calculations with exact-exchange density-functional theory: An analysis of valence-band photoemission for compound semiconductors, New J. Phys. 7~(1) (2005) 126.
\newblock \href {https://doi.org/10.1088/1367-2630/7/1/126} {\path{doi:10.1088/1367-2630/7/1/126}}.

\bibitem{vanschilfgaarde_quasiparticle_2006}
M.~{van~Schilfgaarde}, T.~Kotani, S.~Faleev, Quasiparticle {{Self-Consistent}} {{{\emph{GW}}}} {{Theory}}, Phys. Rev. Lett. 96~(22) (2006) 226402.
\newblock \href {https://doi.org/10.1103/PhysRevLett.96.226402} {\path{doi:10.1103/PhysRevLett.96.226402}}.

\bibitem{vanschilfgaarde_adequacy_2006}
M.~{van Schilfgaarde}, T.~Kotani, S.~V. Faleev, Adequacy of approximations in {{{\emph{GW}}}} theory, Phys. Rev. B 74~(24) (2006) 245125.
\newblock \href {https://doi.org/10.1103/PhysRevB.74.245125} {\path{doi:10.1103/PhysRevB.74.245125}}.

\bibitem{fuchs_quasiparticle_2007}
F.~Fuchs, J.~Furthm{\"u}ller, F.~Bechstedt, M.~Shishkin, G.~Kresse, Quasiparticle band structure based on a generalized {{Kohn-Sham}} scheme, Phys. Rev. B 76~(11) (2007) 115109.
\newblock \href {https://doi.org/10.1103/PhysRevB.76.115109} {\path{doi:10.1103/PhysRevB.76.115109}}.

\bibitem{shishkin_selfconsistent_2007}
M.~Shishkin, G.~Kresse, Self-consistent {{{\emph{GW}}}} calculations for semiconductors and insulators, Phys. Rev. B 75~(23) (2007) 235102.
\newblock \href {https://doi.org/10.1103/PhysRevB.75.235102} {\path{doi:10.1103/PhysRevB.75.235102}}.

\bibitem{schleife_bandstructure_2009}
A.~Schleife, F.~Fuchs, C.~R{\"o}dl, J.~Furthm{\"u}ller, F.~Bechstedt, Band-structure and optical-transition parameters of wurtzite {{MgO}}, {{ZnO}}, and {{CdO}} from quasiparticle calculations, Phys. Status Solidi B 246~(9) (2009) 2150--2153.
\newblock \href {https://doi.org/10.1002/pssb.200945204} {\path{doi:10.1002/pssb.200945204}}.

\bibitem{shih_quasiparticle_2010}
B.-C. Shih, Y.~Xue, P.~Zhang, M.~L. Cohen, S.~G. Louie, Quasiparticle {{Band Gap}} of {{ZnO}}: {{High Accuracy}} from the {{Conventional}} {{{\emph{G}}}}{\emph{{\textsubscript{0}}}}{{{\emph{W}}}}{\emph{{\textsubscript{0}}}} {{Approach}}, Phys. Rev. Lett. 105~(14) (2010) 146401.
\newblock \href {https://doi.org/10.1103/PhysRevLett.105.146401} {\path{doi:10.1103/PhysRevLett.105.146401}}.

\bibitem{friedrich_band_2011}
C.~Friedrich, M.~C. M{\"u}ller, S.~Bl{\"u}gel, Band convergence and linearization error correction of all-electron {{{\emph{GW}}}} calculations: {{The}} extreme case of zinc oxide, Phys. Rev. B 83~(8) (2011) 081101.
\newblock \href {https://doi.org/10.1103/PhysRevB.83.081101} {\path{doi:10.1103/PhysRevB.83.081101}}.

\bibitem{stankovski_g0w0_2011}
M.~Stankovski, G.~Antonius, D.~Waroquiers, A.~Miglio, H.~Dixit, K.~Sankaran, M.~Giantomassi, X.~Gonze, M.~C{\^o}t{\'e}, G.-M. Rignanese, {\emph{G}}{\emph{{\textsubscript{0}}}}{{{\emph{W}}}}{\emph{{\textsubscript{0}}}} band gap of {{ZnO}}: {{Effects}} of plasmon-pole models, Phys. Rev. B 84~(24) (2011) 241201.
\newblock \href {https://doi.org/10.1103/PhysRevB.84.241201} {\path{doi:10.1103/PhysRevB.84.241201}}.

\bibitem{friedrich_hybrid_2012}
C.~Friedrich, M.~Betzinger, M.~Schlipf, S.~Bl{\"u}gel, A.~Schindlmayr, Hybrid functionals and {{{\emph{GW}}}} approximation in the {{FLAPW}} method, J. Phys.: Condens. Matter 24~(29) (2012) 293201.
\newblock \href {https://doi.org/10.1088/0953-8984/24/29/293201} {\path{doi:10.1088/0953-8984/24/29/293201}}.

\bibitem{samsonidze_insights_2014}
G.~Samsonidze, C.-H. Park, B.~Kozinsky, Insights and challenges of applying the {{{\emph{GW}}}} method to transition metal oxides, J. Phys.: Condens. Matter 26~(47) (2014) 475501.
\newblock \href {https://doi.org/10.1088/0953-8984/26/47/475501} {\path{doi:10.1088/0953-8984/26/47/475501}}.

\bibitem{wiktor_comprehensive_2017}
J.~Wiktor, I.~Reshetnyak, F.~Ambrosio, A.~Pasquarello, Comprehensive modeling of the band gap and absorption spectrum of {{BiVO}}{\textsubscript{4}}, Phys. Rev. Mater. 1~(2) (2017) 022401.
\newblock \href {https://doi.org/10.1103/PhysRevMaterials.1.022401} {\path{doi:10.1103/PhysRevMaterials.1.022401}}.

\bibitem{golze_gw_2019}
D.~Golze, M.~Dvorak, P.~Rinke, The {{GW Compendium}}: {{A Practical Guide}} to {{Theoretical Photoemission Spectroscopy}}, Front. Chem. 7 (2019).
\newblock \href {https://doi.org/10.3389/fchem.2019.00377} {\path{doi:10.3389/fchem.2019.00377}}.

\bibitem{rangel_reproducibility_2020}
T.~Rangel, M.~Del~Ben, D.~Varsano, G.~Antonius, F.~Bruneval, F.~H. Da~Jornada, M.~J. Van~Setten, O.~K. Orhan, D.~D. O'Regan, A.~Canning, A.~Ferretti, A.~Marini, G.-M. Rignanese, J.~Deslippe, S.~G. Louie, J.~B. Neaton, Reproducibility in {{{\emph{G}}}}{\emph{{\textsubscript{0}}}}{{{\emph{W}}}}{\emph{{\textsubscript{0}}}} calculations for solids, Computer Physics Communications 255 (2020) 107242.
\newblock \href {https://doi.org/10.1016/j.cpc.2020.107242} {\path{doi:10.1016/j.cpc.2020.107242}}.

\bibitem{vurgaftman_band_2001}
I.~Vurgaftman, J.~R. Meyer, L.~R. {Ram-Mohan}, Band parameters for {{III}}--{{V}} compound semiconductors and their alloys, Journal of Applied Physics 89~(11) (2001) 5815--5875.
\newblock \href {https://doi.org/10.1063/1.1368156} {\path{doi:10.1063/1.1368156}}.

\bibitem{madelung_semiconductors_2004}
O.~Madelung, Semiconductors: {{Data Handbook}}, Springer Berlin Heidelberg, Berlin, Heidelberg, 2004.
\newblock \href {https://doi.org/10.1007/978-3-642-18865-7} {\path{doi:10.1007/978-3-642-18865-7}}.

\bibitem{kondo_structural_2008}
S.~Kondo, K.~Tateishi, N.~Ishizawa, Structural {{Evolution}} of {{Corundum}} at {{High Temperatures}}, Jpn. J. Appl. Phys. 47~(1S) (2008) 616.
\newblock \href {https://doi.org/10.1143/JJAP.47.616} {\path{doi:10.1143/JJAP.47.616}}.

\bibitem{mizoguchi_strong_2004}
H.~Mizoguchi, P.~M. Woodward, C.-H. Park, D.~A. Keszler, Strong {{Near-Infrared Luminescence}} in {{BaSnO}}{\textsubscript{3}}, J. Am. Chem. Soc. 126~(31) (2004) 9796--9800.
\newblock \href {https://doi.org/10.1021/ja048866i} {\path{doi:10.1021/ja048866i}}.

\bibitem{sleight_crystal_1979}
A.~W. Sleight, H.~y.~Chen, A.~Ferretti, D.~E. Cox, Crystal growth and structure of {{BiVO}}{\textsubscript{4}}, Materials Research Bulletin 14~(12) (1979) 1571--1581.
\newblock \href {https://doi.org/10.1016/0025-5408(72)90227-9} {\path{doi:10.1016/0025-5408(72)90227-9}}.

\bibitem{werner_highpressure_1982}
A.~Werner, H.~D. Hochheimer, High-pressure x-ray study of {{Cu}}{\textsubscript{2}}{{O}} and {{Ag}}{\textsubscript{2}}{{O}}, Phys. Rev. B 25~(9) (1982) 5929--5934.
\newblock \href {https://doi.org/10.1103/PhysRevB.25.5929} {\path{doi:10.1103/PhysRevB.25.5929}}.

\bibitem{recker_directional_1988}
K.~Recker, F.~Wallrafen, K.~Dupr{\'e}, Directional solidification of the {{LiF-LiBaF}}{\textsubscript{3}} eutectic, Naturwissenschaften 75~(3) (1988) 156--157.
\newblock \href {https://doi.org/10.1007/BF00405314} {\path{doi:10.1007/BF00405314}}.

\bibitem{sugiyama_crystal_1991}
K.~Sugiyama, Y.~Tak{\'e}uchi, {The crystal structure of rutile as a function of temperature up to 1600{$^\circ$} C}, Z. F{\"u}r Krist. - Cryst. Mater. 194~(1-4) (1991) 305--314.
\newblock \href {https://doi.org/10.1524/zkri.1991.194.14.305} {\path{doi:10.1524/zkri.1991.194.14.305}}.

\bibitem{garcia-martinez_microstructural_1993}
O.~{Garc{\'i}a-Mart{\'i}nez}, R.~M. Rojas, E.~Vila, J.~L.~M. {de Vidales}, Microstructural characterization of nanocrystals of {{ZnO}} and {{CuO}} obtained from basic salts, Solid State Ionics 63--65 (1993) 442--449.
\newblock \href {https://doi.org/10.1016/0167-2738(93)90142-P} {\path{doi:10.1016/0167-2738(93)90142-P}}.

\bibitem{cardona_isotope_2005}
M.~Cardona, M.~L.~W. Thewalt, Isotope effects on the optical spectra of semiconductors, Rev. Mod. Phys. 77~(4) (2005) 1173--1224.
\newblock \href {https://doi.org/10.1103/RevModPhys.77.1173} {\path{doi:10.1103/RevModPhys.77.1173}}.

\bibitem{miglio_predominance_2020}
A.~Miglio, V.~{Brousseau-Couture}, E.~Godbout, G.~Antonius, Y.-H. Chan, S.~G. Louie, M.~C{\^o}t{\'e}, M.~Giantomassi, X.~Gonze, Predominance of non-adiabatic effects in zero-point renormalization of the electronic band gap, npj Comput Mater 6~(1) (2020) 1--8.
\newblock \href {https://doi.org/10.1038/s41524-020-00434-z} {\path{doi:10.1038/s41524-020-00434-z}}.

\bibitem{strossner_dependence_1986}
K.~Str{\"o}ssner, S.~Ves, C.~Koo~Kim, M.~Cardona, Dependence of the direct and indirect gap of {{AlSb}} on hydrostatic pressure, Phys. Rev. B 33~(6) (1986) 4044--4053.
\newblock \href {https://doi.org/10.1103/PhysRevB.33.4044} {\path{doi:10.1103/PhysRevB.33.4044}}.

\bibitem{clark_intrinsic_1997}
C.~D. Clark, P.~J. Dean, P.~V. Harris, W.~C. Price, Intrinsic edge absorption in diamond, Proc. R. Soc. Lond. Ser. Math. Phys. Sci. 277~(1370) (1997) 312--329.
\newblock \href {https://doi.org/10.1098/rspa.1964.0025} {\path{doi:10.1098/rspa.1964.0025}}.

\bibitem{piacentini_thermoreflectance_1976}
M.~Piacentini, D.~W. Lynch, C.~G. Olson, Thermoreflectance of {{LiF}} between 12 and 30 {{eV}}, Phys. Rev. B 13~(12) (1976) 5530--5543.
\newblock \href {https://doi.org/10.1103/PhysRevB.13.5530} {\path{doi:10.1103/PhysRevB.13.5530}}.

\bibitem{nery_quasiparticles_2018}
J.~P. Nery, P.~B. Allen, G.~Antonius, L.~Reining, A.~Miglio, X.~Gonze, Quasiparticles and phonon satellites in spectral functions of semiconductors and insulators: {{Cumulants}} applied to the full first-principles theory and the {{Fr{\"o}hlich}} polaron, Phys. Rev. B 97~(11) (2018) 115145.
\newblock \href {https://doi.org/10.1103/PhysRevB.97.115145} {\path{doi:10.1103/PhysRevB.97.115145}}.

\bibitem{whited_exciton_1969}
R.~C. Whited, W.~C. Walker, Exciton {{Spectra}} of {{CaO}} and {{MgO}}, Phys. Rev. Lett. 22~(26) (1969) 1428--1430.
\newblock \href {https://doi.org/10.1103/PhysRevLett.22.1428} {\path{doi:10.1103/PhysRevLett.22.1428}}.

\bibitem{french_interband_1994}
R.~H. French, D.~J. Jones, S.~Loughin, Interband {{Electronic Structure}} of {$\alpha$}-{{Alumina}} up to 2167 {{K}}, J. Am. Ceram. Soc. 77~(2) (1994) 412--422.
\newblock \href {https://doi.org/10.1111/j.1151-2916.1994.tb07009.x} {\path{doi:10.1111/j.1151-2916.1994.tb07009.x}}.

\bibitem{park_applicability_2022}
J.~Park, W.~A. Saidi, B.~Chorpening, Y.~Duan, Applicability of {{Allen}}--{{Heine}}--{{Cardona Theory}} on {{MO}}{\textsubscript{x}} {{Metal Oxides}} and {{ABO}}{\textsubscript{3}} {{Perovskites}}: {{Toward High-Temperature Optoelectronic Applications}}, Chem. Mater. 34~(13) (2022) 6108--6115.
\newblock \href {https://doi.org/10.1021/acs.chemmater.2c01281} {\path{doi:10.1021/acs.chemmater.2c01281}}.

\bibitem{aggoune_consistent_2022}
W.~Aggoune, A.~Eljarrat, D.~Nabok, K.~Irmscher, M.~Zupancic, Z.~Galazka, M.~Albrecht, C.~Koch, C.~Draxl, A consistent picture of excitations in cubic {{BaSnO}}{\textsubscript{3}} revealed by combining theory and experiment, Commun Mater 3~(1) (2022) 1--10.
\newblock \href {https://doi.org/10.1038/s43246-022-00234-6} {\path{doi:10.1038/s43246-022-00234-6}}.

\bibitem{lushchik_multiplication_1994}
C.~Lushchik, E.~Feldbach, A.~Frorip, M.~Kirm, A.~Lushchik, A.~Maaroos, I.~Martinson, Multiplication of electronic excitations in {{CaO}} and {{YAlO}}{\textsubscript{3}} crystals with free and self-trapped excitons, J. Phys.: Condens. Matter 6~(50) (1994) 11177.
\newblock \href {https://doi.org/10.1088/0953-8984/6/50/025} {\path{doi:10.1088/0953-8984/6/50/025}}.

\bibitem{komornicki_structural_2004}
S.~Komornicki, M.~Radecka, P.~Soba{\'s}, Structural, electrical and optical properties of {{TiO}}{\textsubscript{2}}--{{WO}}{\textsubscript{3}} polycrystalline ceramics, Materials Research Bulletin 39~(13) (2004) 2007--2017.
\newblock \href {https://doi.org/10.1016/j.materresbull.2004.07.017} {\path{doi:10.1016/j.materresbull.2004.07.017}}.

\bibitem{wu_firstprinciples_2018}
Y.-N. Wu, W.~A. Saidi, P.~Ohodnicki, B.~Chorpening, Y.~Duan, First-{{Principles Investigations}} of the {{Temperature Dependence}} of {{Electronic Structure}} and {{Optical Properties}} of {{Rutile TiO}}{\textsubscript{2}}, J. Phys. Chem. C 122~(39) (2018) 22642--22649.
\newblock \href {https://doi.org/10.1021/acs.jpcc.8b06941} {\path{doi:10.1021/acs.jpcc.8b06941}}.

\bibitem{cooper_indirect_2015}
J.~K. Cooper, S.~Gul, F.~M. Toma, L.~Chen, Y.-S. Liu, J.~Guo, J.~W. Ager, J.~Yano, I.~D. Sharp, Indirect {{Bandgap}} and {{Optical Properties}} of {{Monoclinic Bismuth Vanadate}}, J. Phys. Chem. C 119~(6) (2015) 2969--2974.
\newblock \href {https://doi.org/10.1021/jp512169w} {\path{doi:10.1021/jp512169w}}.

\bibitem{wang_temperature_2003}
L.~Wang, N.~C. Giles, Temperature dependence of the free-exciton transition energy in zinc oxide by photoluminescence excitation spectroscopy, Journal of Applied Physics 94~(2) (2003) 973--978.
\newblock \href {https://doi.org/10.1063/1.1586977} {\path{doi:10.1063/1.1586977}}.

\end{thebibliography}

\end{document}